\documentclass[twocolumn,showpacs,prb,superscriptaddress]{revtex4-1}
\usepackage{graphicx}
\usepackage{color}\usepackage{enumerate}
\usepackage{amsmath}\usepackage{amssymb}\usepackage{stmaryrd}
\usepackage{esint}
\usepackage{multirow}
\usepackage{float}
\usepackage{longtable,booktabs}
\usepackage{subfigure}
\usepackage{dsfont}
\usepackage{amssymb}
\usepackage{comment}
\usepackage[pdftex,colorlinks=true,linkcolor=blue,citecolor=blue]{hyperref}
\usepackage[normalem]{ulem}
\usepackage{mathtools}

\DeclarePairedDelimiter\ket{\lvert}{\rangle}
\DeclarePairedDelimiterX\braket[2]{\langle}{\rangle}{#1 \delimsize\vert #2}

\usepackage{tikz}

\begin{document}
\title{Lieb Schultz Mattis-Type Theorems and Other Non-perturbative Results for Strongly Correlated Systems with Conserved Dipole Moments}
\author{Oleg Dubinkin}
\email[]{olegd2@illinois.edu}
\author{Julian May-Mann}
\email[]{maymann2@illinois.edu}
\author{Taylor L. Hughes}
\email[]{hughest@illinois.edu}
\affiliation{Department of Physics and Institute for Condensed Matter Theory, University of Illinois at Urbana-Champaign, IL 61801, USA}
\begin{abstract}
Non-perturbative constraints on many body physics--such as the famous Lieb-Schultz-Mattis theorem--are valuable tools for studying strongly correlated systems. To this end, we present a number of non-perturbative results that constrain the low-energy physics of systems having conserved dipole moments. We find that for these systems, a unique translationally invariant gapped ground state is only possible if the polarization of the system is integer. Furthermore, if a lattice system also has $U(1)$ subsystem charge conservation symmetry, a unique gapped ground state is only possible if the particle filling along these subsystems is integer. We also apply these methods to spin systems, and determine criteria for the existence of a new type of magnetic response plateau in the presence of a non-uniform magnetic field. Finally, we formulate a version of Luttinger's theorem for 1D systems consisting of dipoles.

\end{abstract}

\maketitle
\section{Introduction}\label{sec:intro}
Strongly correlated quantum systems are often the source of striking phenomena, yet they remain one of the most challenging to analyze. If perturbative approaches fail, one's only recourse is numerical simulation unless non-perturbative methods or results can be applied. One of the most celebrated non-perturbative results is the Lieb-Schultz-Mattis (LSM) theorem, which from very little information, i.e., the number of spin-$1/2$ degrees of freedom per unit cell, can draw conclusions about the low-energy properties of 1D spin chains\cite{lieb1961}. This result has had wide-ranging applicability in quantum systems, and has been extended to higher dimensions\cite{oshikawa2000A,hastings2004,hastings2005,nachtergaele2007}. 

A key feature of the proof of the LSM theorem is a twist operator that slowly rotates the spins across a spin-chain. In the work of Oshikawa\cite{oshikawa2000A}, which applied and generalized the LSM result to lattice systems with conserved particle number,  a related twist operator is used
\begin{equation}
U_{X}=\exp\left[ \frac{2\pi i\hat{X}}{L_x}\right],
\label{eqn:twist_op}
\end{equation}\noindent where $\hat{X}=\sum_{{\bf{x}}} x \hat{n}_{\bf{x}}$ is the many-body position operator. Indeed, Oshikawa and collaborators also used this operator to provide a non-perturbative understanding of Luttinger's theorem\cite{luttinger1960, oshikawa1997}, determine the Fermi surface properties of the Kondo lattice\cite{oshikawa1997,oshikawa2000B, ueda1994, otsuki2009}, and more recently to calculate filling-enforced constraints on the quantum Hall conductivity in lattice systems\cite{lu2017filling}. Remarkably, this operator $U_{X}$ has had parallel uses in the theory of electronic polarization where it was introduced by Resta\cite{resta1998}. In this context, the complex phase of the ground state expectation value of $U_{X}$ is determined by the electronic polarization\cite{resta1998}, and the magnitude is determined by the electron localization length\cite{resta1999,aligia1999,souza2000}.

In recent work, new twist operators have been proposed whose ground state expectation values can be used to calculate higher multipole moments\cite{wheeler2018many,gil2018,dubinkinmaymannhughes2019,you2019}. The simplest multipole generalization of $U_{X}$ is the operator
\begin{equation}
U_{XY}=\exp\left[\frac{2\pi i\widehat{XY}}{L_x L_y}\right]
\label{eqn:multipole_twist}
\end{equation}\noindent where $\widehat{XY}=\sum_{{\bf{x}}}x y \hat{n}_{\bf{x}}$ is the many-body quadrupole operator. In light of this development it is natural to use these multipole operators to try to derive non-perturbative results analogous to the previous work on the LSM theorem\cite{oshikawa2000A,lu2017lieb}, Luttinger's theorem\cite{oshikawa1997,oshikawa2000B}, magnetization plateaus\cite{affleck1989, oshikawa1997mag}, and filling-enforced Hall conductivity constraints\cite{lu2017filling}. In this article we focus on higher multipole generalizations of some of these results, derived through the application of operators related to $U_{XY}$ in each context. Our goal is to recast the original results that apply to particles/charges to apply to dipoles. Notably, we study generalizations of the LSM theorem for systems that conserve dipole moments (Section \ref{sec:dip_LSM}), and then apply these results to study magnetization (gradient) plateaus in spin systems (Section \ref{sec:plateau}), and an extension of Luttinger's theorem to dipole conserving systems (Section \ref{sec:luttinger}).  We also extend our results on the LSM-type theorems to systems with $U(1)$ subsystem symmetry, e.g., symmetries enforcing charge conservation along rows or columns in 2D\cite{batista2005,you2018subsystem}. These results are applicable to some fracton systems, and it is possible they may eventually be adapted in some form to systems with broken subsystem symmetry\cite{you2018higher,you2019} (and broken microscopic dipole conservation), e.g., higher order multipole band insulators of fermions or bosons\cite{benalcazar2017quantized,you2018higher,dubinkin2018higher}, though we leave those developments to future work.

\section{LSM-Type Theorems for Dipole Conserving Systems}
\label{sec:dip_LSM}

Let us briefly recount the concept behind Oshikawa's proof\cite{oshikawa2000A} of the LSM theorem that we wish to generalize. His starting point was a system with $U(1)$ particle conservation, and he utilized a twist operator $U_X$ that achieves two different things simultaneously: when applied to the ground state of a periodic insulating system, it extracts the charge polarization of the ground state, \emph{and} it performs a large gauge transformation on this state, i.e., it effectively inserts or removes a single unit of magnetic flux through the periodic loop running along the $\hat{x}$-direction. To see how this works, consider a nearest-neighbor lattice model of free fermions in 1D with $N_x$ sites along $\hat{x}:$ $H=-t\sum_{j=1}^{N_x}(c^{\dagger}_{j+1}c^{\phantom{\dagger}}_j+h.c.)$. Applying $U_X$ to the Hamiltonian modifies every fermionic hopping term by attaching a phase factor of $\text{e}^{2\pi i/N_x},$ which is equivalent to introducing an external gauge field $A_x$ with a $2\pi$ circulation along the loop going around the periodic $\hat{x}$-direction. If $U_X$ is applied to the ground state $\vert\Psi_0\rangle$ of this free-fermion system, for example, then, in the thermodynamic limit, a set of filled bands will return back to themselves up to a phase proportional to the polarization, while a partially filled band will change momentum and be orthogonal to the original state. This is a very simple application of Oshikawa's results\cite{oshikawa2000A} that indicates that partially filled bands cannot support an excitation gap since the energy of the state $U_x\vert\Psi_0\rangle$ will nominally approach that of $\vert \Psi_0\rangle$ itself in the thermodynamic limit, and the two states are orthogonal if there are partially filled bands because they have different momentum quantum numbers.

\subsection{Dipole Conserving Systems}
\label{sec:dip_LSM_A}
To make progress toward an LSM-type theorem for systems with charge and dipole conservation, let us consider a system defined on a rectangular periodic $L_x \times L_y=N_x a\times N_y a,$ lattice where $a$ is the lattice constant. We will work with Hamiltonians $H$ that are translationally invariant and conserve both global particle and dipole number. Hence, the Hamiltonian is invariant when the charged operators are changed by constant phase transformations $\text{e}^{i\alpha},$ and phase transformations with linear coordinate dependence $\text{e}^{i{\boldsymbol{\alpha}}\cdot{\bf{x}}},$ respectively. The latter condition also automatically implies that $H$ commutes with the twist operators $U_X$ and $U_Y$, i.e., the total polarization is a fixed quantum number for eigenstates of $H.$ We note that it has been shown\cite{pretko18gauge} that systems with dipole conservation can be coupled to a rank-2 gauge field $A_{ij}$. The rank-2 gauge field transforms as $A_{ij} \rightarrow A_{ij} + \partial_{i}\partial_j \lambda$ under a gauge transformation, where $\lambda$ is a generic function.

Similar to Oshikawa's work, we will be considering the action of generalized $U_{XY}$ twist operators on the ground states of insulating systems: $U_{XY}(\alpha)=\text{e}^{i\alpha\widehat{XY}}$. For $\alpha=\frac{2\pi}{L_x L_y},$ and for systems with open boundaries, this operator was shown to be successful in extracting the quadrupolar polarization\cite{wheeler2018many,kang2018many,dubinkinmaymannhughes2019} and, when applied to systems that explicitly conserve dipole moment, it introduces a constant rank-2 gauge field $A_{xy}$ across the lattice. However, if periodic boundary conditions are introduced, $U_{XY}(\frac{2\pi}{L_x L_y})$ exhibits problematic behavior. For instance, while the complex \emph{phase} of its ground-state expectation value $\langle U_{XY}(\frac{2\pi}{L_x L_y})\rangle_0,$ computed for the ground states of free-fermion tight-binding models, correctly captures the quadrupolar polarization, the absolute value of the expectation value vanishes when the thermodynamic limit is taken because of fluctuations of the dipole moments. Even if we mitigate this issue by restricting ourselves to Hamiltonians with manifest dipole conservation, as we shall do in this article, the dipole-conserving terms in the Hamiltonian that cross the periodic boundary pick up an additional phase factor under the action of $U_{XY}(\frac{2\pi}{L_x L_y}).$ To see this, one can act on dipole-conserving terms in the Hamiltonian, e.g., ring-exchange terms
\begin{equation}
\begin{split}
    U^{-1}_{XY}(\alpha)\Big(c^{\dagger}_{x,y}c^{\phantom{\dagger}}_{x+a,y}&c^{\dagger}_{x+a,y+a}c^{\phantom{\dagger}}_{x,y+a}\Big)U_{XY}(\alpha)\nonumber\\=&\text{e}^{i\alpha}c^{\dagger}_{x,y}c^{\phantom{\dagger}}_{x+a,y}c^{\dagger}_{x+a,y+a}c^{\phantom{\dagger}}_{x,y+a}.\nonumber
\end{split}
\end{equation} For terms that cross a periodic boundary additional phase factors are generated, e.g., $\text{e}^{i\alpha L_x a}$ or $\text{e}^{i\alpha L_y a}.$ Indeed, one can check that in order to have $U_{XY}(\alpha)$ insert a constant rank-2 gauge field $A_{xy}$ for a system with periodic boundary conditions then $\alpha L_x a =2\pi\mathbb{Z}$ and $\alpha L_y a=2\pi \mathbb{Z}.$ Hence, in order to be consistent with periodic boundary conditions, we will choose $\alpha=\frac{2\pi}{a^2 \gcd(N_x, N_y)}$ where $a$ is the lattice constant in the $x$ and $y$ directions.

 To proceed from this setup, let us consider adiabatically turning on a constant gauge field configuration of  $A_{xy}$ over a time $T$ having the form 
\begin{equation}
    A_{xy} = \frac{2\pi }{a^2\gcd(N_x,N_y)}\frac{t}{T}.
\end{equation}  Let us label the Hamiltonian as a function of time as $H(t),$ and its instantaneous ground state as $\ket{\Psi(t)}$. Since the initial system is translationally invariant, $\ket{\Psi(0)}$ is an eigenstate of the many-body translation operators $T_x$ and $T_y$ that send each particle coordinate $(x,y)$ to $(x+a,y)$ or $(x,y+a)$ respectively. We will take the $T_x$ eigenvalue to be $\text{e}^{iP_{x0}}$ and the $T_y$ eigenvalue to be $\text{e}^{iP_{y0}}$. Since $H(t)$ is translationally invariant at all times,  $\ket{\Psi(t)}$ will remain an eigenstate of $T_x$ and $T_y$ with the same eigenvalues $\text{e}^{iP_{x0}}$ and $e^{iP_{y0}}$ at all times. Similarly, since $H(t)$ commutes with both $U_X$ and $U_Y$ at all times, $\ket{\Psi(t)}$ will also remain an eigenstate of $U_X$ and $U_Y$ with eigenvalues $\text{e}^{2\pi i X_0/L_x}$ and $\text{e}^{ 2\pi i Y_0/L_y}$ at all times. At $t = T$, we have $A_{xy} = \frac{2\pi }{a^2\gcd(N_x,N_y)},$ which is equivalent to a (large) gauge transformation\cite{you2019} that can be removed by applying the many-body twist operator $U_{XY}(\alpha_g)$ where $\alpha_g = \frac{2\pi}{a^2\gcd(N_x,N_y)}$. As a result
\begin{equation}
H(0) = U_{XY}^{-1}(\alpha_g)H(T)U_{XY}(\alpha_g),
 \end{equation}
and so $U_{XY}^{-1}(\alpha_g)\ket{\Psi(T)}$ is also a ground state of $H(0)$. 

The next key step for an LSM-type theorem is to determine if $U_{XY}^{-1}(\alpha_g)\ket{\Psi(T)}$ and $\ket{\Psi(0)}$ are orthogonal. To do this, we will compare the eigenvalues of the translation operator for $U_{XY}^{-1}(\alpha_g)\ket{\Psi(T)}$  and $\ket{\Psi(0)}$. A simple calculation shows that 
\begin{equation}
\begin{split}
    T_x U^{-1}_{XY}&(\alpha_g)\ket{\Psi(T)}\\
    &= \exp\left[\frac{2\pi i \hat{Y}}{a\gcd(N_x,N_y)}\right] e^{i P_{x0}} U^{-1}_{XY}(\alpha_g) \ket{\Psi(T)}.
\end{split}
\end{equation}
We already know that $T_x\ket{\Psi(0)}=\text{e}^{iP_{x0}}\ket{\Psi(0)},$ so the ground state is unique only if $\exp\left[\frac{2\pi i \hat{Y}}{a\gcd(N_x,N_y)}\right]\ket{\Psi(T)} = \ket{\Psi(T)}$. If we define $n=L_y/(a\gcd(N_x,N_y))$, then $\exp\left[\frac{2\pi i \hat{Y}}{a \gcd(N_x,N_y)}\right] = (U_Y)^n$, and 
\begin{equation}
    \exp\left[\frac{2\pi i \hat{Y}}{a\gcd(N_x,N_y)}\right]\ket{\Psi(T)} = \text{e}^{2\pi i n Y_0/L_y}\ket{\Psi(T)},
\end{equation}
where we have used that $U_Y \ket{\Psi(T)} = \text{e}^{2\pi i Y_0/L_y}\ket{\Psi(T)}$. Since $(U_Y)^{N_y} = 1$, $\text{e}^{2\pi i Y_0/L_y}$ must be an $N_y^{\text{th}}$ root of unity, and $Y_0/L_y \equiv  
\frac{1}{2\pi i} \log(e^{2\pi i Y_0/L_y})$ must be a rational number. We can define $Y_0/L_y = p/q$, where $p$ and $q$ are co-prime. So $U^{-1}_{XY}(\alpha_g)\ket{\Psi(T)}$ is an eigenstate of $T_x$ with eigenvalue $\exp(iP_{x0} + 2\pi i n p/q)$. If $n$ and $q$ are co-prime $U^{-1}_{XY}(\alpha_g)\ket{\Psi(T)}$ must be orthogonal to $\ket{\Psi(0)}$, and thus the system is either gapless, or there must be at least $q$ degenerate ground states if the system is gapped.

This statement relies on $n$ and $q$ being co-prime, but if $n$ is an integer multiple of $q$ nothing can be said about the degeneracy of the ground state. A similar issue was remarked upon in Oshikawa's proof of the LSM theorem\cite{oshikawa2000A}. Here, we can avoid this issue by requiring that the thermodynamic limit is taken such that $N_x = N_y=N$. Since ground state properties should be independent of how the thermodynamic limit is taken, we can assume that the the thermodynamic limit is taken in this way. From this we can conclude that the ground state of the system is unique only if $\text{e}^{2\pi i Y_0/Na} = 1$. Using the same logic, i.e., by acting with the translation operator $T_y,$ we can also conclude that the ground state is unique only if $\text{e}^{2\pi i X_0/Na} = 1$ as well. In other words, for the ground state to be unique, we must require that the components of the polarization in both directions vanish up to a polarization quantum. The bulk of a locally electrically neutral system must carry a uniform polarization, which allows us to relate a microscopic dipole moment stretching between the pair of neighboring unit cells to a macroscopic polarization of the system picked up by the phase of the unitary operator $\text{e}^{2\pi i \hat{X}_j/N_ja}$. Therefore, we see that the pair of conditions $\text{e}^{2\pi i Y_0/Na}=\text{e}^{2\pi i X_0/Na}=1$ is equivalent to requiring that the microscopic dipole moments stretching between every pair of neighboring unit cells must be an integer times $ea$. Thus we conclude that filling factor for $x$ and $y$ dipoles must be an integer, analogous to the requirement that the charge filling factor be integer in the conventional LSM theorem.

\subsection{Subsystem Symmetric Systems}
\label{sec:subs_LSM}
We can construct a stronger version of the above dipole LSM theorem for Hamiltonians with dipole conservation arising from $U(1)$ subsystem symmetry. Let us first provide a brief background discussion on subsystem symmetries. To give an example, consider a 2-dimensional $L_x\times L_y$ rectangular lattice. The subsystem symmetry operator corresponding to  $U(1)$ charge conservation along a single column with $x=x_0$ is given by:
\begin{equation}
    U_{0,x_0}(\alpha) = \exp\left(i\alpha \sum^{L_y}_{y=1} \hat{n}_{x_0,y}\right).    
\end{equation}
Such operators rotate the phase of all electrons along a single column in the lattice. In other words, $U_{0,x_0}$ can be thought of as a restriction of the global $U(1)$ symmetry operator $U_0(\alpha)=\exp\left(i\alpha \sum^{L_x,L_y}_{x,y=1} \hat{n}_{x,y}\right)$ to a particular subsystem. Similarly, we can define subsystem symmetries $U_{0,y_0}$ that impose charge conservation along every single row $y_0$  of the lattice. For the purposes of our work, by an $n$-dimensional subsystem in a $d$-dimensional Bravais lattice, we will understand an $n$-dimensional lattice subspace spanned by any $n$ linearly independent lattice basis vectors. 
Now, taking a collection of such ``parallel" subspaces that cover the entire lattice, we can impose $U(1)$ charge conservation along \emph{each} of the subspaces individually. This restriction leads to a conservation of \emph{all} multipole moments\cite{gromov2018multipole} in a $(d-n)$-dimensional subspace orthogonal to these subsystems. 

Coming back to our two-dimensional lattice example, take a collection of parallel lattice lines that cover the whole lattice. For concreteness, let us take a collection of lattice rows that are parallel to $\hat{x}$. 
Imposing charge conservation along each row is equivalent to fixing the total charge at each position along $\hat{y},$ which is the normal vector to these subsystems. 
Thus, for any arbitrary function $f(y)$, the conservation of the quantity
\begin{equation}
    Q=\sum_{x,y} f(y)q(x,y),
\end{equation}
where $q(x,y)$ is the charge at a site with coordinates $(x,y)$, is guaranteed by the $U(1)$ subsystem charge conservation.
For example, by taking $f(y)=y^m$, we can see that all multipole moments along $\hat{y}$, such as the $P_y$ component of the dipole moment, the $Q_{yy}$ component of the quadrupole moment, etc., are conserved.
Similarly, imposing charge conservation along every row of sites parallel to the $\hat{x}$-axis in a 3D lattice leads to the conservation of all multipole moments in the $yz$-plane, e.g., $P_y$, $P_z$, $Q_{yy}$, $Q_{yz}$, $Q_{zz}$, etc.

Furthermore, we can impose subsystem charge conservation along two different families of subsystems simultaneously, e.g., rows and columns in 2D. For a two-dimensional lattice this leads to a conservation of both components of the dipole moment as well as all higher multipole moments diagonal in either $x$ or $y$ coordinates.
However, these subsystem symmetries do not guarantee conservation of all multipole moments with components along the subsystem. 
In particular, conservation of the off-diagonal component of the quadrupole moment $Q_{xy}$ is not achieved by imposing charge conservation along rows parallel to $\hat{x}$ and columns parallel to $\hat{y}$. Instead, one would have to impose charge conservation along rows of sites that are perpendicular to the $xy$-plane. 
In 2D, for example, this translates into requiring charge conservation at each individual site of the lattice, which trivially leads to a conservation of \emph{all} multipole moments of such system. For the majority of this article we will be focused on the simple case of 1D and 2D systems having subsystem symmetries along the rows and/or columns.

After that brief discussion let us develop an LSM-type theorem for systems with subsystem symmetry. First, let us consider a periodic $L_x\times L_y$ rectangular lattice with a Hamiltonian $H$ that conserves $U(1)$ charge along every row and every column of the lattice, i.e., we have $[H,U_{0,x_0}(\alpha)]=[H,U_{0,y_0}(\alpha)]=0$ for every $x_0$, $y_0,$ and $\alpha$. Particle and dipole conserving Hamiltonians satisfying these criteria can be built from subsystem-symmetric ring-exchange type terms\cite{dubinkinmaymannhughes2019,xu2007,pretko2019emergent}.

Now, consider the following twist operator acting along a single column of sites with fixed coordinate $x=x_0$:
\begin{equation}
    U_{Y,x_0}=\exp\left(\frac{2\pi i}{L_y}\sum_{y}y\hat{n}_{x_0,y}\right).
    \label{eqn:1d_multipole_twist}
\end{equation}
For a periodic lattice and a dipole-conserving Hamiltonian built from local ring-exchange type interactions (say, of fermions) such as
\begin{equation}
    \begin{split}
    H=-t_{\square}\sum_{\textbf{r}}\left(c^\dagger_{\textbf{r}+\hat{x}}c^{\ }_{\textbf{r}} c^{\dagger}_{\textbf{r}+\hat{y}}c^{\ }_{\textbf{r}+\hat{x}+\hat{y}}+h.c.\right),
    \end{split}
    \label{eqn:ring-exchange}
\end{equation}
we can calculate the energy difference between the ground state $\vert\Psi_0\rangle$, and a twisted ``variational" state $U_{Y,x_0}\vert\Psi_0\rangle$: \begin{equation}
\begin{split}
    &\langle \Psi_0|U_{Y,x_0}^{-1} H U_{Y,x_0} - H|\Psi_0 \rangle\\
    =&-t_{\square}\bigg[\left(\text{e}^{2\pi i/N_y}-1\right)\sum_{y}\langle c^\dagger_{x_0+a,y}c_{x_0,y}c^\dagger_{x_0,y+a}c_{x_0+a,y+a}\rangle\\
    &+\left(\text{e}^{-2\pi i/N_y}-1\right)\sum_{y}\langle c^\dagger_{x_0,y}c_{x_0-a,y}c^\dagger_{x_0-a,y+a}c_{x_0,y+a}\rangle\bigg]\\
    &+h.c.
    \end{split}
    \label{eqn:ham_difference}
\end{equation}
We can follow analogous calculations to Refs. \onlinecite{lieb1961,oshikawa1997} by expanding the exponents in the powers of $1/N_y,$ and assuming that the ground state preserves at least one of the reflection symmetries. The constant term in the Taylor expansion of the exponential factors immediately cancels, but we we need to check that the next order term also vanishes so that the first non-vanishing term is actually the second-order term from the exponential, which would imply that the energy difference is ultimately of the order $O(1/N_y)$.
To see this, after expanding the exponential, consider a pair of plaquettes related by mirror $\hat{M}_y: y\to -y$ and consider the following sum of two of their ring-exchange terms from Eq. \ref{eqn:ham_difference}:
\begin{equation}
\begin{split}
\frac{2\pi i}{N_y}&\langle c^\dagger_{x_0+a,y}c_{x_0,y}c^\dagger_{x_0,y+a}c_{x_0+a,y+a}\rangle\\
&-\frac{2\pi i}{N_y}\langle c^\dagger_{x_0+a,-y}c_{x_0,-y}c^\dagger_{x_0,-y-a} c_{x_0+a,-y-a}\rangle.
\end{split}
\label{eqn:sum_reflection}
\end{equation}
We note that the second term, having an opposite sign, comes from the hermitian conjugate part of the overall Hamiltonian (\ref{eqn:ring-exchange}). If the ground state is an eigenstate of the reflection symmetry, $\hat{M}_y\ket{\Psi_0}=\pm\ket{\Psi_0}$, we can rewrite the second term as:
\begin{equation}
\begin{split}
    \langle&\Psi_0|c^\dagger_{x_0+a,-y}c_{x_0,-y}c^\dagger_{x_0,-y-a} c_{x_0+a,-y-a}|\Psi_0\rangle \\
    &=\langle\Psi_0|\hat{M}^{-1}_y c^\dagger_{x_0+a,-y}c_{x_0,-y}c^\dagger_{x_0,-y-a} c_{x_0+a,-y-a}\hat{M}_y|\Psi_0\rangle\\
    &=\langle\Psi_0| c^\dagger_{x_0+a,y}c_{x_0,y}c^\dagger_{x_0,y+a}c_{x_0+a,y+a}|\Psi_0\rangle,
    \end{split}
\end{equation}
and so we see that the sum in Eq. \ref{eqn:sum_reflection} exactly vanishes. The same analysis is applicable to every other pair of plaquettes that are related by $\hat{M}_y,$ and so we conclude that the first non-vanishing term in Eq. \ref{eqn:ham_difference} is of the order $O(1/N_y)$.
Therefore, in the thermodynamic limit, where $N_y\to\infty$, the state $\ket{\tilde{\Psi}_0}=U_{Y,x_0}\ket{\Psi_0}$ has  either exactly the same energy as the ground state $\ket{\Psi_0},$ or it is an excited state with an infinitesimally small energy. 

Using this result, if we can now show that $\ket{\Psi_0}$ and $\ket{\tilde{\Psi}_0}$ are orthogonal, then the system must necessarily be gapless (or at least have ground state degeneracy). 
To this end, let us assume that the ground state does not spontaneously break the translational symmetry, so $\ket{\Psi_0}$ will be an eigenstate of the translation operator $T_x$. Using $T_x\ket{\Psi_0}=\text{e}^{i P_{x0}}\ket{\Psi_0}$ it follows that
\begin{equation}
    \begin{split}
        T_x\ket{\tilde{\Psi}_0}&=T_xU_{Y,x_0}T_x^{-1}T_x\ket{\Psi_0}\\
        &=\text{e}^{iP_{x0}+\frac{2\pi i}{L_y}\sum_{y=a}^{L_y}y(\hat{n}_{x_0+a,y}-\hat{n}_{x_0,y})}\ket{\tilde{\Psi}_0}.
    \end{split}
\end{equation}
With the subsystem charge conservation in place, we can introduce a subsystem polarization associated with a subsystem $\mathfrak{s}$ as follows:
\begin{equation}
    \mathcal{P}^j_{\mathfrak{s}}=\frac{e}{2\pi}\text{Im}\log\langle\Psi_0|U_{j,\mathfrak{s}}|\Psi_0\rangle
    \label{eqn:subsystem_pol}
\end{equation}
where
\begin{equation}
    U_{j,\mathfrak{s}}=\exp\left(\frac{2\pi i}{L_j}\sum_{\textbf{r}\in \mathfrak{s}}x_j \hat{n}_{\textbf{r}}\right),
\end{equation} and where $j=x,y.$
Now the condition on the ground state momentum shift to be an integer times $2\pi$ can be understood as a condition on the polarizations of two neighboring subsystems:
\begin{equation}
\begin{split}
    \text{e}^{\frac{2\pi i}{L_y}\sum_{y=a}^{L_y}y(\hat{n}_{x_0+a,y}-\hat{n}_{x_0,y})}\ket{\tilde{\Psi}_0}&=\ket{\tilde{\Psi}_0}\\
    \Leftrightarrow \tfrac{1}{e}\left(\mathcal{P}^y_{x=x_0+a}\right.&-\left.\mathcal{P}^y_{x=x_0}\right)\in\mathbb{Z},
    \end{split}
    \label{eqn:subs_pol_diff}
\end{equation}
where $\mathcal{P}^y_{x=x_0}$ is the polarization (\ref{eqn:subsystem_pol}) along the column with fixed coordinate $x=x_0$. In other words, the $\hat{x}$ lattice derivative of the $y$-polarization must be an integer. In general, since we can translate by any number of lattice constants in the $x$-direction, each column must have $\mathcal{P}^{y}_{x=x_0}$ that differ from each other at most by an integer. While this may seem at odds with translation symmetry, the integer ambiguity in the polarization for periodic systems implies that the subsystem polarizations can differ by an integer (or more generally can differ by a polarization quantum) and still maintain translation symmetry. Now, recall the statement from the previous subsection, that, in order for dipole-conserving system on an $N\times N$ lattice to support a gap, we require the total polarization to vanish. Combining this result with Eq. \ref{eqn:subs_pol_diff} we find that the polarization of each subsystem can take only the following possible of values if the system is gapped:
\begin{equation}
    \mathcal{P}^j_{\mathfrak{s}}=\frac{n^j_{\mathfrak{s}}}{N}e\mod e,\  n^j_{\mathfrak{s}}\in\mathbb{Z}.
\end{equation}

Next, we can derive a condition using the fact that the ground state $\ket{\Psi_0}$ will also be an eigenstate of the translation operator $T_y$: $T_y\ket{\Psi_0}=\text{e}^{i P_{y0}}\ket{\Psi_0}.$ Hence for the twisted state $\ket{\tilde{\Psi}_0}$ we find:
\begin{equation}
    T_y\ket{\tilde{\Psi}_0}=T_yU_{Y,x_0}T_y^{-1}T_y\ket{\Psi_0}=\text{e}^{iP_{y0}+\frac{2\pi i}{N_y}\sum_{y=1}^{N_y}\hat{n}_{x_0,y}}\ket{\tilde{\Psi}_0}.
\end{equation}
Thus, $\ket{\tilde{\Psi}_0}$ is an eigenstate of $T_y$ with eigenvalue $\exp(i P_{y0} + 2\pi i\nu^{x_0}),$ where $\nu^{x_0}$ is the filling factor of the subsystem $x=x_0$. Hence, the states $\ket{\tilde{\Psi}_0}$ and $\ket{\Psi_0}$ are orthogonal unless each column has integer filling $\nu^{x_0}$. So the system will either have degenerate ground states or gapless excitations if $\nu^{x_0} \notin \mathbb{Z}$ for any column. We can go back and repeat this analysis for subsystem symmetries at fixed $y_0.$ We will find that $T_x$ will require that the subsystem filling factors $\nu^{y_0}$ must all be integers for the ground state to be gapped/non-degenerate. Analogously, $T_y$ will require that the subsystem polarizations $\mathcal{P}^{x}_{y=y_0}$ must all be the same up to an integer (polarization quantum) for the ground state to be gapped/non-degenerate. Although we have chosen a specific Hamiltonian and imposed reflection symmetry to illustrate that the twisted state has low energy, we expect that the result is much more general\cite{tasaki2018lieb}, just as the original LSM result is.

One consequence of these results is that the regular LSM theorem derived for systems with global $U(1)$ charge conservation symmetry\cite{oshikawa2000A} must be satisfied along every subsystem individually. In the original case, the LSM theorem states that for the ground state to be unique, the particle number per unit cell calculated across the whole lattice must be an integer. Our arguments invoking subsystem symmetry and translation can be straightforwardly applied to $d$-dimensional Hamiltonians conserving total $U(1)$ charge across $n$-dimensional subsystems. We can use this logic to formulate a generalization of the LSM theorem to classes of Hamiltonians that posses not only global $U(1)$ symmetry but also subsystem $U(1)$ symmetries: 

\emph{Consider a $d$-dimensional periodic lattice with a short-range Hamiltonian that respects $U(1)$ subsystem charge conservation across a family of parallel $n$-dimensional subspaces. Assume that the ground state does not spontaneously break translational symmetry along the subsystem.
For an arbitrary lattice  there is at least one low-energy state degenerate, or infinitesimally close in energy, to the ground state, if the particle number per unit cell $\nu^{\mathfrak{s}}$ in any particular $n$-dimensional subsystem $\mathfrak{s}$ is not an integer.} Furthermore, the low-energy state has a crystal momentum in the $j$-direction (associated to a lattice translation operator $T_j$ that leaves the subsystem invariant) differing from the ground state by an amount $\Delta P_j =2\pi\frac{V^{\mathfrak{s}}}{L_j}\nu^{\mathfrak{s}}$, where $V^{\mathfrak{s}}$ is the volume of an individual subsystem, and $L_j$ is the lattice size along the $\hat{x}_j$.

To provide a proof for this general statement, we can use a suitably adapted version of the argument in Ref. \onlinecite{lu2017filling}. Let us consider a general (bosonic or fermionic) Hamiltonian $H$ defined on a periodic $d$-dimensional lattice that conserves the total $U(1)$ charge across an $n$-dimensional subspace spanned by a collection of $n$ linearly independent primitive lattice vectors $\{\vec{a}_1,...,\vec{a}_n\},$ and their integer linear combinations. We will label this subspace as $\mathfrak{s}$. There are a family of subsystems ``parallel" to $\mathfrak{s}$, but for now let us focus on this one subsystem. The $U(1)$ symmetry operator for this subsystem is given by
\begin{equation}
    U_{0,\mathfrak{s}}(\alpha) = \exp \left(i \alpha \sum_{\textbf{r} \in \mathfrak{s}} \hat{n}_{\textbf{r}}\right),
\end{equation}
where $\textbf{r}$ is a lattice vector. 

The most general lowest-order Hamiltonian that acts on $\mathfrak{s}$ while commuting with $U_{0,\mathfrak{s}}(\alpha)$ takes the following form
\begin{equation}
     H=\sum_{\textbf{r}_1,\textbf{r}_2,j} J_{\textbf{r}_1,\textbf{r}_2,j}\ c^\dagger_{\textbf{r}_1}c_{\textbf{r}_2}\hat{\mathcal{O}}_j + h.c.
     \label{eqn:Hamiltonian}
\end{equation}
where $J_{\textbf{r}_1,\textbf{r}_2,j}$ is a set of coupling constants, $\textbf{r}_1,\textbf{r}_2 \in \mathfrak{s},$ and $\hat{\mathcal{O}}_j$ are products of particle creation and annihilation operators that have no support on $\mathfrak{s}.$ The $\hat{\mathcal{O}}_j$ may be further constrained by subsystem symmetries for other subspaces $\mathfrak{s}'$, but we will treat them as arbitrary. If we want to gauge the subsystem $U(1)$ symmetry it requires the introduction of an $(n+1)$-dimensional vector-potential $A^{\mathfrak{s}}_i$, $i=0...n$ associated with subsystem $\mathfrak{s}$. Individual terms in the Hamiltonian Eq. (\ref{eqn:Hamiltonian}) couple to the lattice gauge field, which modifies the overall Hamiltonian with a subsystem Peierls factor:
\begin{equation}
     H(A^{\mathfrak{s}})=\sum_{\textbf{r}_1,\textbf{r}_2,j}\left( J_{\textbf{r}_1,\textbf{r}_2,j}\ \text{e}^{iA^{\mathfrak{s}}_{\textbf{r}_1,\textbf{r}_2}}c^\dagger_{\textbf{r}_1}c_{\textbf{r}_2}\hat{\mathcal{O}}_j +h.c.\right).
\end{equation} 

Now we will invoke the momentum counting argument for the subsystem $\mathfrak{s}$ following the presentation of Ref. \onlinecite{lu2017filling}, with the main difference being that the gauge field is now associated with subsystem $U(1)$ charge conservation instead of a regular global $U(1)$ symmetry. Let us define subsystem magnetic flux quantum insertion operators $\mathcal{F}^{\mathfrak{s}}_j$, $j=1...n$ that adiabatically, over time period $T$, evolves the gauge field $A^{\mathfrak{s}}$ between two configurations that differ by a large gauge transformation performed along the direction spanned by the $\vec{a}_j$-th primitive vector.
Adiabaticity of the process is required so that the gap never closes during the time evolution and that, by starting with a ground state $\ket{\Psi_0}$, we are guaranteed to end up with a (possibly different) state $\ket{\tilde{\Psi}_0}$ which lies in the ground state subspace, after the time evolution is finished.
Let us define the time evolution of the Hamiltonian as:
\begin{equation}
    H_j(t)=H\left(A^{\mathfrak{s}}_j=\frac{2\pi}{N_j a}\frac{t}{T}\right),
\end{equation}
where $N_j$ is the lattice period along the primitive vector $\vec{a}_j$ having lattice constant $a$, and the subsystem lattice gauge field is picked to be uniform in space. 
The corresponding time evolution operator is then given by
\begin{equation}
    \mathcal{F}^{\mathfrak{s}}_j=\mathcal{T}\exp\left(-i\int_0^T dt\ H_j(t)\right),
\end{equation}
where $\mathcal{T}$ denotes time-ordering. 
The final Hamiltonian $H_j(T)$ differs from the initial one $H_j(0)$ by a large gauge transformation, which can be implemented by the following operator: 
\begin{equation}
    U_{X_j,\mathfrak{s}}=\exp\left(\frac{2\pi i}{L_j}\sum_{\textbf{r}\in \mathfrak{s}}r_j \hat{n}_{\textbf{r}}\right),
\end{equation}
and one can directly verify that:
\begin{equation}
    H_j(T)=U_{X_j,\mathfrak{s}}^{-1}H_j(0)U_{X_j,\mathfrak{s}}.
\end{equation} We can now define the subsystem flux insertion-removal operator that leaves the Hamiltonian invariant:
\begin{equation}
    \tilde{\mathcal{F}}^{\mathfrak{s}}_j\equiv U_{X_j,\mathfrak{s}}^{-1}\mathcal{F}^{\mathfrak{s}}_j.
\end{equation}

Importantly, the initial ground state $\ket{\Psi_0}$, in general, might be different from the state $\ket{\tilde{\Psi}_0}\equiv\tilde{\mathcal{F}}^{\mathfrak{s}}_j\ket{\Psi_0}$ obtained after the subsystem flux insertion and removal procedure. To see this explicitly, let us consider the action of the translation operator $T_{j}$ on the states $\ket{\Psi_0}$ and $\ket{\tilde{\Psi}_0}$ (we note that these translations preserve the subsystem $\mathfrak{s}$). Provided that the translational invariance is not spontaneously broken, we must have:
\begin{equation}
    T_j\ket{\Psi_0}=\text{e}^{iP_{j0}}\ket{\Psi_0}
\end{equation}
where $P_{j0}$ is the many-body momentum along $\vec{a}_j$, and is a good quantum number modulo $2\pi$. To see how $T_j$ acts on $\ket{\tilde{\Psi}_0}$ we first note that, since the subsystem gauge field evolution implemented by $\mathcal{F}^{\mathfrak{s}}_j$ does not break translational invariance, we have:
\begin{equation}
    [\mathcal{F}^{\mathfrak{s}}_j,T_j]=0.
\end{equation}
However, $T_j$ acts non-trivially on the flux removal unitary: 
\begin{equation}
    T_j U_{X_j,\mathfrak{s}} T_j^{-1}=\text{e}^{-\frac{2\pi i}{N_j}\sum_{\textbf{r}\in\mathfrak{s}}\hat{n}_{\textbf{r}}}U_{X_j,\mathfrak{s}}.
\end{equation}
Therefore, the $T_j$ eigenvalue of the final state is:
\begin{equation}
\begin{split}
    T_j\ket{\tilde{\Psi}_0}&=\mathcal{F}^{\mathfrak{s}}_jT_j U_{X_j,\mathfrak{s}}T_j^{-1}T_j\ket{\Psi_0}\\
    &=\exp\left(iP_{j0}+2\pi i\frac{V^{\mathfrak{s}}}{N_j}\nu^{\mathfrak{s}}\right)\ket{\tilde{\Psi}_0},
\end{split}
\end{equation}
where $V^{\mathfrak{s}}$ is the total number of unit cells inside the subsystem $\mathfrak{s},$ and $\nu^{\mathfrak{s}}$ is the filling fraction of said subsystem $\nu^{\mathfrak{s}}=\frac{\mathcal{N}^{\mathfrak{s}}}{V^{\mathfrak{s}}}$, where $\mathcal{N}^{\mathfrak{s}}$ is the total particle number in $\mathfrak{s}$. We thus see that the momentum along $\vec{a}_j$ of the two states $\ket{\Psi_0}$ and $\ket{\tilde{\Psi}_0}$ differs by:
\begin{equation}
    \Delta P_{j}=2\pi \frac{V^{\mathfrak{s}}}{N_j}\nu^{\mathfrak{s}}.
\end{equation}
Hence, whenever $\frac{V^{\mathfrak{s}}}{N_j}\nu^{\mathfrak{s}}$ is not an integer, the two states $\ket{\Psi_0}$ and $\ket{\tilde{\Psi_0}}$ must be orthogonal to each other. 

The same argument can be applied to the flux insertion-removal procedure for every direction along the subsystem $\mathfrak{s}$.
Therefore, for the ground state to be unique, we must require that:
\begin{equation}
    \frac{V^{\mathfrak{s}}}{N_j}\nu^{\mathfrak{s}}\in\mathbb{Z},\ j=1...n.
    \label{eqn:conditions}
\end{equation}
Since each $N_j$ is a divisor of $V^{\mathfrak{s}}$ the ratios $V^{\mathfrak{s}}/N_j$ are all integers. However, for this set of conditions to be satisfied for arbitrary lattice sizes, we must require $\nu^{\mathfrak{s}}$ itself to be integer. For example, in the case where all $N_i$ are co-prime with each other, the only way to satisfy all of the conditions (\ref{eqn:conditions}) is to require that
\begin{equation}
    \nu^{\mathfrak{s}}\in\mathbb{Z}.
\end{equation}
We thereby arrive at the theorem stated at the end of the Section \ref{sec:subs_LSM}. 

We can also provide a lower bound on the ground state degeneracy in the case when $\nu^{\mathfrak{s}}\notin\mathbb{Z}$. Assuming that $\frac{V^{\mathfrak{s}}}{N_j}\nu^{\mathfrak{s}}=\frac{p^{\mathfrak{s}}_j}{q^{\mathfrak{s}}_j}$, where the pair of integers $p^{\mathfrak{s}}_j$ and $q^{\mathfrak{s}}_j$ are co-prime for all $j=1...n$:
\begin{equation}
    GSD\geq \prod_{j=1}^n q^{\mathfrak{s}}_j.
\end{equation}
If we now consider subsystem charge conservation along $m$ subsystems $\mathfrak{s}_i$, $i=1,\ldots m,$ which are all translationally invariant in the $j$-th direction (e.g., parallel rows), we can combine different twist operators $U_{X_j,\mathfrak{s_i}}$, $i=1,...,m$ to generate low-lying states where the many-body momentum is shifted when compared to the ground state by:
\begin{equation}
    \Delta P_{j}=2\pi \frac{V^{\mathfrak{s}}}{N_j}\sum_{i=1}^m n_i\nu^{\mathfrak{s_i}},
\end{equation}
where $n_i$ are integers and $V^{\mathfrak{s}}$ is the total number of unit cells in every subsystem.
The total number of inequivalent values (mod $2\pi$) that $\Delta P_{j}$ can take is the least common multiple of corresponding $q^{\mathfrak{s}_i}_j$ integers.
Thus, the ground state degeneracy associated with translations $T_j$ is bounded below by
\begin{equation}
    GSD_j\ge \text{lcm}\left(q^{\mathfrak{s}_1}_j,q^{\mathfrak{s}_2}_j,...,q^{\mathfrak{s}_m}_j\right).
\end{equation}
Hence, the total ground state degeneracy for a $d$-dimensional lattice with $m$ subsystem symmetries is bounded below by:
\begin{equation}
    GSD\ge \prod_{j=1}^d \text{lcm}\left(q^{\mathfrak{s}_1}_j,q^{\mathfrak{s}_2}_j,...,q^{\mathfrak{s}_m}_j\right)
\end{equation}
where we set $q^{\mathfrak{s}_i}_j=1$ if a particular subsystem $\mathfrak{s}_i$ is not left invariant under the action of $T_j$. 

To give an example, imagine a two-dimensional lattice where, on top of the regular global $U(1)$ symmetry we impose independent subsystem $U(1)$ symmetries along two rows $\mathfrak{r}_1$ and $\mathfrak{r}_2$ defined by the equations $y=y_1$ and $y=y_2$ respectively.
Both $\mathfrak{r}_1$ and $\mathfrak{r}_2$ are invariant under the action of $T_x$ operator.
As an example, let us now choose the fillings of these two rows to satisfy:
\begin{equation}
    \frac{V^{\mathfrak{r}_1}}{N_x}\nu^{\mathfrak{r}_1}=\frac12\equiv\frac{1}{q_x^{\mathfrak{r}_1}},\quad \frac{V^{\mathfrak{r}_2}}{N_x}\nu^{\mathfrak{r}_2}=\frac14\equiv\frac{1}{q_x^{\mathfrak{r}_2}}.
\end{equation}
Applying the unitary operator $U_{X,\mathfrak{r}_1}$, we can conclude that there are at least two states in the ground state subspace which are eigenstates of the translation operator $T_x$ with eigenvalues $\text{e}^{iP_{x0}}$ and $\text{e}^{iP_{x0}+i\pi}$. Similarly, applying unitary $U_{X,\mathfrak{r}_2}$ we find at least four translationally invariant states with eigenvalues $\text{e}^{iP_{x0}}$, $\text{e}^{iP_{x0}+i\pi/2}$, $\text{e}^{iP_{x0}+i\pi}$, and $\text{e}^{iP_{x0}+i3\pi/2}.$ Since two of these values have already appeared when we used $U_{X,\mathfrak{r}_1}$, and we cannot easily distinguish two twisted states with the same momentum,  we conclude that the total ground state degeneracy is at least $4=\text{lcm}(q_x^{\mathfrak{r}_1},q_x^{\mathfrak{r}_2})$.

As another example let us consider a pair of subsystem symmetries imposed along two columns $\mathfrak{c}_1$ and $\mathfrak{c}_2$ which are defined by the equations $x=x_1$ and $x=x_2$ with fillings
\begin{equation}
    \frac{V^{\mathfrak{c}_1}}{N_y}\nu^{\mathfrak{c}_1}=\frac12\equiv\frac{1}{q_y^{\mathfrak{c}_1}},\quad \frac{V^{\mathfrak{c}_2}}{N_y}\nu^{\mathfrak{c}_2}=\frac13\equiv\frac{1}{q_y^{\mathfrak{c}_2}}.
\end{equation}
Analogous to the previous paragraph, these subsystems are invariant under $T_y,$ and acting on the ground state with unitary operators $U_{Y,\mathfrak{c}_1}$ and $U_{Y,\mathfrak{c}_2}$ generates low-lying states with translation eigenvalues $\text{e}^{iP_{y0}}$ and $\text{e}^{iP_{y0}+i\pi}$ for the first operator, and $\text{e}^{iP_{y0}}$, $\text{e}^{iP_{y0}+i2\pi/3}$, and $\text{e}^{iP_{y0}+i4\pi/3}$ for the second one.
Additionally, we can combine $U_{Y,\mathfrak{c}_1}$ with $U_{Y,\mathfrak{c}_2}$ to obtain a low-energy state with the translation eigenvalue equal to $\text{e}^{iP_{y0}+i\pi/3}$ and $\text{e}^{iP_{y0}+i5\pi/3}$.
Thus, we conclude that the total number of low-lying states that can be generated by the column twist operators is $6=\text{lcm}(q_y^{\mathfrak{c}_1},q_y^{\mathfrak{c}_2})$. 

For a two-dimensional system that is translationally invariant simultaneously along both $\hat{x}$ and $\hat{y}$, we expect the fillings of all rows to be the same if the ground state does not spontaneously break translation symmetry, therefore $q_x=q^{\mathfrak{r}_1}_{x}=q^{\mathfrak{r}_2}_{x}=...$, and so the ground state degeneracy associated with translations $T_x$ is
\begin{equation}
    GSD_x\ge \text{lcm}\left(q^{\mathfrak{r}_1}_x,q^{\mathfrak{r}_2}_x,...\right)=q_x.
\end{equation}
Similarly, the fillings of all columns must also be the same giving us $q_y=q^{\mathfrak{c}_1}_{y}=q^{\mathfrak{c}_2}_{y}=...$, leading to the ground state degeneracy associated with translations along $\hat{y}$ to be
\begin{equation}
    GSD_y\ge \text{lcm}\left(q^{\mathfrak{c}_1}_y,q^{\mathfrak{c}_2}_y,...\right)=q_y.
\end{equation}
The total ground state degeneracy of such system is then bounded from below by the product of the two factors:
\begin{equation}
    GSD\ge q_x q_y.
\end{equation}

\section{Application: Plateaus in Magnetization and Magnetization Gradients}\label{sec:plateau}
Let us take these concepts and apply them to spin systems with an aim toward making physical predictions. In the work of Oshikawa, Yamanaka, and Affleck\cite{oshikawa1997mag}, bosonic spin counterparts of the twist operators (\ref{eqn:twist_op}) were successfully used to derive conditions for the appearance of magnetization plateaus as a function of applied external magnetic field in, e.g., spin chains. Here we will define bosonic spin counterparts of the multipole twist operators Eqs. (\ref{eqn:multipole_twist}), (\ref{eqn:1d_multipole_twist}) to derive conditions in spin ladder systems for the appearance of plateaus of the \emph{gradient} of magnetization as a function of an applied magnetic field \emph{gradient} placed across the ladder.  Explicitly, we imagine tuning the magnetic field so that the system is on a conventional magnetization plateau, and then test how the gradient of the magnetization responds to a magnetic field gradient around a uniform background field. We will want to distinguish between two cases: (i) the system has a non-constant magnetization gradient response,  or (ii) the system exhibits a plateau in the magnetization gradient. These physical phenomena are closely related to the recent work on dipole insulators\cite{dubinkinmaymannhughes2019}. In the language of Ref. \onlinecite{dubinkinmaymannhughes2019}, since the system is tuned to a conventional magnetization plateau, the analogous charge system would be a charge insulator. However, case (i) would be a charge insulator but a dipole metal, while (ii) would be both a charge insulator and a dipole insulator.

To illustrate these two possibilities we will consider systems with axial spin rotation symmetry along subspaces, i.e., models with $U(1)$ subsystem symmetry corresponding to a conservation of the total $S^z$ along each subspace. 
A unitary operator corresponding to a $U(1)$ subsystem symmetry associated to a particular subsystem $\mathfrak{s}$ reads: 
\begin{equation}
    U_{0,\mathfrak{s}}(\alpha) = \exp\left(i\alpha \sum_{\textbf{r}\in\mathfrak{s}} \hat{S}^{z}_{\textbf{r}}\right).    
\end{equation}
This operator rotates all spins belonging to $\mathfrak{s}$ around the $z$-axis by the same amount. The corresponding conserved quantity is the total $S^z$ magnetization on $\mathfrak{s}$.

\subsection{One-dimensional Spin Ladder Model}
As an explicit test system let us first consider a two-leg spin-$S$ ladder that is stretched along the $\hat{x}$-axis with periodic boundary conditions in the $x$-direction. We will also enforce two $U(1)$ subsystem symmetries, one of which implies conservation of the total magnetization on the top leg (which we label with `$\uparrow$'), and the other which implies conservation of the total magnetization of the bottom leg  (which we label with `$\downarrow$'). Let us assume that the system's ground state does not break translational symmetry, and has a fixed total magnetization for some range of values of an external magnetic field $h_0 < B^z < h_1$, i.e., the state of the system is at a magnetization plateau. Applying the conventional magnetization plateau argument\cite{oshikawa1997mag} to a two-leg ladder spin system we find the magnetization \emph{per spin} in a two-leg ladder of the size $L_x\times L_y = N_x a\times 2a$:
\begin{equation}
M^z\equiv\frac{1}{2N_x}\sum_{x=1}^{N_x}(S^z_{x,\uparrow}+S^z_{x,\downarrow})  \label{eq:magperspin}  
\end{equation} 
takes half-integer values, i.e., $M^z=0,\pm \frac12,\pm 1,$ etc.

To preserve the subsystem symmetries we can build a Hamiltonian from spin ring-exchange terms, e.g., nearest neighbor ring exchanges:
\begin{equation}
    H=J_{\square}\sum_{x}\left(S^+_{x,\uparrow}S^-_{x,\downarrow}S^+_{x+1,\downarrow}S^-_{x+1,\uparrow}+h.c.\right).
    \label{eqn:ham_bos_ring-exchange}
\end{equation}
Similar to the ring-exchange model studied in the previous section, where such terms tunneled charge dipoles, here they can be interpreted as tunneling terms for magnetic quadrupole moments (spins are already magnetic dipoles so separating opposite spins by a distance to create a ``dipole of spins" creates a magnetic quadrupole). Now, consider the following unitary twist operator acting along one of the legs of the ladder:
\begin{equation}
    U_{X,\uparrow}=\exp\left(\frac{2\pi i}{L_x}\sum_{x}x S^z_{x,\uparrow}\right).
    \label{eqn:magnetic_unitary}
\end{equation} Under the action of the operator $U_{X,\uparrow}$ each term in $H$ is modified as:
\begin{equation}
    \begin{split}
        U^{-1}_{X,\uparrow}S^+_{x,\uparrow}S^-_{x,\downarrow}&S^+_{x+1,\downarrow}S^-_{x+1,\uparrow}U_{X,\uparrow}\\
        &=\text{e}^{\frac{2\pi i}{N_x}}S^+_{x,\uparrow}S^-_{x,\downarrow}S^+_{x+1,\downarrow}S^-_{x+1,\uparrow}.
    \end{split}
\end{equation}
Therefore, we can show that for the ground state $\ket{\Psi_0},$ which we assume preserves translation and reflection symmetry along $\hat{x},$ we have:
\begin{equation}
    \langle\Psi_0 |U_{X,\uparrow}^{-1} H U_{X,\uparrow} - H |\Psi_0\rangle=O\left(\frac{1}{N_x}\right).
\end{equation}
And so, similar to the previous section, we see that in the thermodynamic limit $N_x\to \infty$ the state $\ket{\tilde{\Psi}_0}=U_{X,\uparrow}\ket{\Psi_0}$ lies in, or infinitesimally near, the ground state subspace.

Now let us check if $\ket{\tilde{\Psi}_0}=U_{X,\uparrow}\ket{\Psi_0}$ is orthogonal to $\vert\Psi_0\rangle.$ Following logic that should now be apparent, we can compute the commutation relation between the translation operator and $U_{X,\uparrow}$ to find:
\begin{equation}
    T_x U_{X,\uparrow} T^{-1}_{x}=U_{X,\uparrow} \text{e}^{2\pi i S^z_{1,\uparrow}-\frac{2\pi i}{N_x}\sum_{x=1}^{N_x}S^z_{x,\uparrow}}.
\end{equation}
Therefore, starting from a ground state $\ket{\Psi_0}$ having a well-defined many-body momentum $T_x\ket{\Psi_{0}}=\text{e}^{iP_{x0}}\ket{\Psi_{0}}$, we find the state $\ket{\tilde{\Psi}_0}$ has the eigenvalue:
\begin{equation}
        \begin{split}
        T_x\ket{\tilde{\Psi}_0}&=T_xU_{X,\uparrow}T_x^{-1}T_x\ket{\Psi_0}\\
        &=\text{e}^{iP_{x0}+2\pi i S^z_{1,\uparrow}-\frac{2\pi i}{N_x}\sum_{x=1}^{N_x}S^z_{x,\uparrow}}\ket{\tilde{\Psi}_0}.
    \end{split}
\end{equation} A notable difference from the previous section is the appearance of the extra term $2\pi S_{1,\uparrow}$ in the phase factor which can be integer or half-integer depending on the spin model of interest. From this analysis we conclude that the two states are orthogonal unless $S_{\uparrow}-m_{\uparrow}\in\mathbb{Z},$ where $m_{\uparrow}=\tfrac{1}{N_x}\sum_{x}S^{z}_{x,\uparrow}.$
We can obtain a similar condition by considering the unitary operator $U_{X,\downarrow}$ that acts on the bottom leg.
Thus, the ground state can be unique only if the spin minus the average magnetization $m^z_{\uparrow/\downarrow}$ of the top row or bottom row of spins are both integers:
\begin{equation}
    S_{\uparrow/\downarrow}-m^z_{\uparrow/\downarrow}\in\mathbb{Z}.
\end{equation}

Let us analyze these conditions in more detail. We can actually make a more physically intuitive statement by noticing that the sum of average magnetizations of both rows must be an integer as well: 
\begin{equation}
    m^z_\uparrow + m^z_\downarrow=2M^z\in \mathbb{Z},
\end{equation}\noindent where we have used the assumption that our system is tuned to a conventional magnetization plateau, and the fact that the magnetization per spin (Eq. \ref{eq:magperspin}) must be a multiple of $1/2$ on the plateau.
We can rewrite magnetizations on both legs as:
\begin{equation}
\begin{split}
    m^z_\downarrow=&2M^z-m^z_\uparrow,\  m^z_\uparrow=n-S,\ n\in\mathbb{Z}.
    \end{split}
\end{equation}
Combining these statements we end up with the following condition for the \emph{magnetization gradient} in the direction transverse to the legs of the ladder: 
\begin{equation}
    \Delta_y m^z \equiv (m^z_\uparrow - m^z_\downarrow)=2(n-S-M^z).
    \label{eq:MGradDef}
\end{equation} In this equation the spin $S$ (in a unit cell on a single leg) is fixed, and since we are tuned to a magnetization plateau $M^z$ is a multiple of $1/2.$ 
Thus, we expect the magnetization gradient to have plateaus at only even or only odd integer values where the parity is determined by  whether the sum of the total spin $S$ and magnetization $M^z$ is integer, or half-integer respectively.  Alternatively, since the total magnetization of the ladder is vanishing, we can recast the magnetization gradient as a magnetic quadrupole moment $Q^{M}_{yz}$ where $z$-oriented magnetizations are separated along the $y$-direction. Our results imply that the system has a plateau of $Q^{M}_{yz}$ as a function of magnetic field gradient. If the total magnetization were on a plateau, but non-vanishing then the conversion to a magnetic quadrupole moment would depend on our choice of coordinate origin. In all of the examples here the total magnetization vanishes so this issue does not arise.

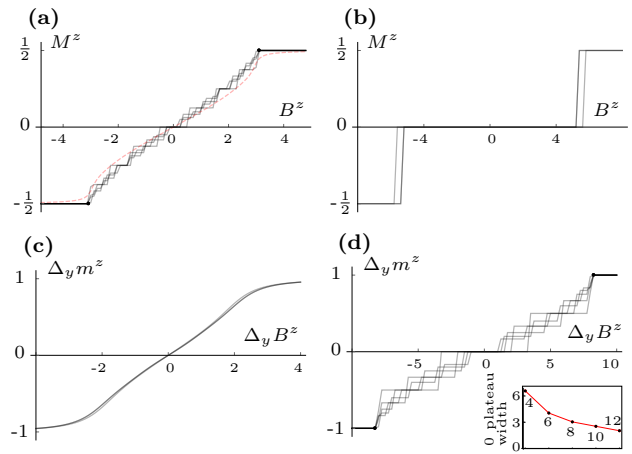
\begin{figure}
\begin{tikzpicture}
    \node[anchor=south west,inner sep=0] at (0,0) {\includegraphics[width=0.21\textwidth]{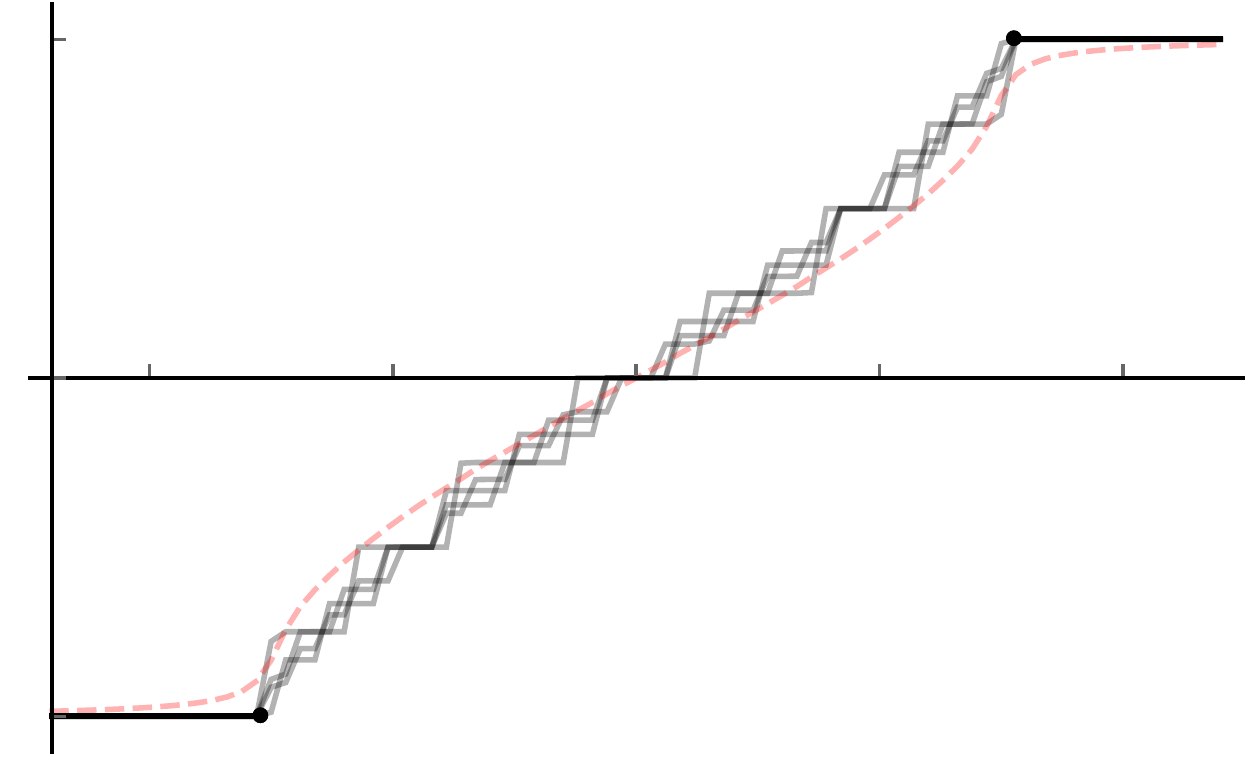}};
    \node at (0.5,2.3) {\scriptsize$M^z$};
    \node at (3.5,1.4) {\scriptsize$ B^z$};
    \node at (-0.05,0.15) {\scriptsize{-$\frac12$}};
    \node at (-0.05,1.2) {\scriptsize{0}};
    \node at (-0.05,2.2) {\scriptsize{$\frac12$}};
    \node at (0.4,1) {\tiny{-4}};
    \node at (1.1,1) {\tiny{-2}};
    \node at (1.9,1) {\tiny{0}};
    \node at (2.65,1) {\tiny{2}};
    \node at (3.4,1) {\tiny{4}};
    \node at (0.2,2.6) {\footnotesize{\textbf{(a)}}};
    
\end{tikzpicture}
\begin{tikzpicture}
    \node[anchor=south west,inner sep=0] at (0,0) {\includegraphics[width=0.21\textwidth]{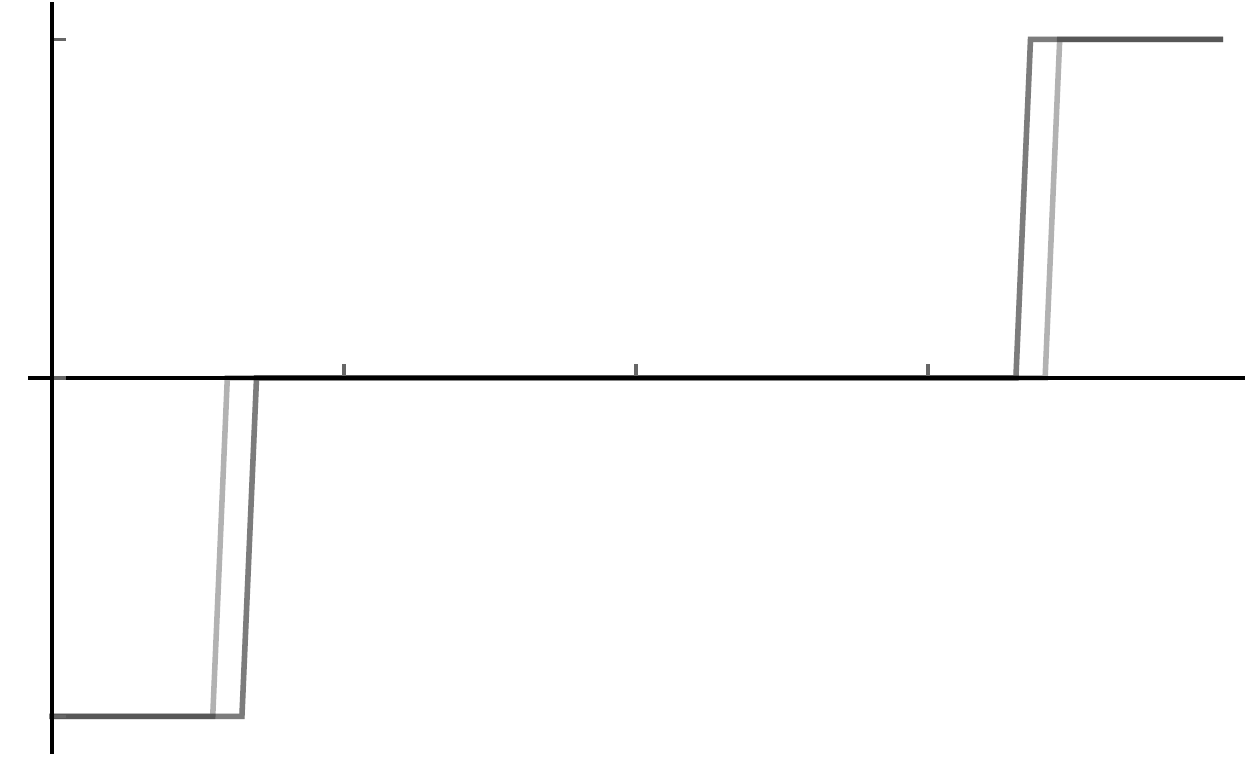}};
    \node at (0.5,2.3) {\scriptsize$M^z$};
    \node at (3.48,1.4) {\scriptsize$ B^z$};
    \node at (-0.05,0.15) {\scriptsize{-$\frac12$}};
    \node at (-0.05,1.2) {\scriptsize{0}};
    \node at (-0.05,2.2) {\scriptsize{$\frac12$}};
    \node at (1,1) {\tiny{-4}};
    \node at (1.95,1) {\tiny{0}};
    \node at (2.8,1) {\tiny{4}};
    \node at (0.2,2.6) {\footnotesize{\textbf{(b)}}};
    
\end{tikzpicture}\\
\begin{tikzpicture}
    \node[anchor=south west,inner sep=0] at (0,0) {\includegraphics[width=0.21\textwidth]{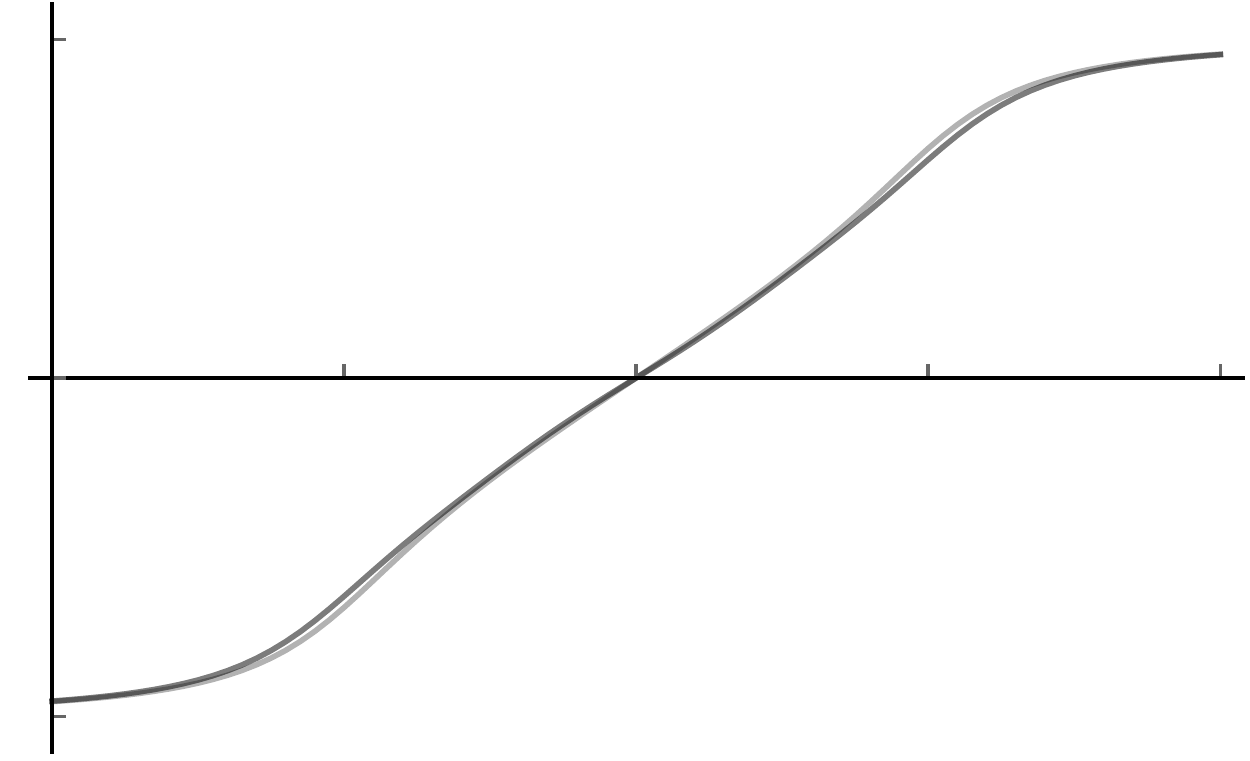}};
    \node at (0.7,2.3) {\scriptsize$\Delta_y m^z$};
    \node at (3.3,1.4) {\scriptsize$\Delta_y B^z$};
    \node at (-0.05,0.15) {\scriptsize{-1}};
    \node at (-0.05,1.2) {\scriptsize{0}};
    \node at (-0.05,2.2) {\scriptsize{1}};
    \node at (1,1) {\tiny{-2}};
    \node at (1.95,1) {\tiny{0}};
    \node at (2.8,1) {\tiny{2}};
    \node at (3.65,1) {\tiny{4}};
    \node at (0.2,2.6) {\footnotesize{\textbf{(c)}}};
    \def\x{1.9}
    \def\y{-1.48}
    \draw[white] (0.43+\x,1.27+\y) -- (1.75+\x,1.27+\y) -- (1.75+\x,2.1+\y) -- (0.43+\x,2.1+\y) -- (0.43+\x,1.27+\y);
\end{tikzpicture}
\begin{tikzpicture}
    \node[anchor=south west,inner sep=0] at (0,0) {\includegraphics[width=0.21\textwidth]{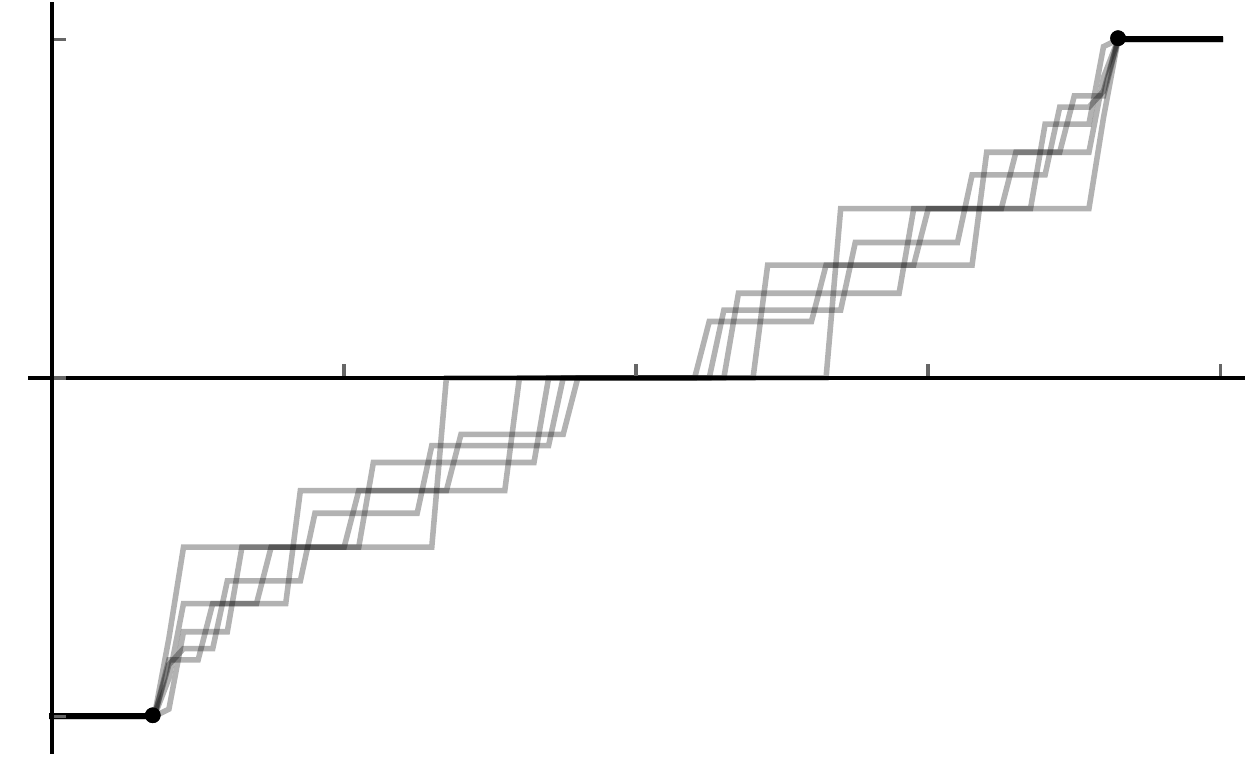}};
    \node at (0.7,2.3) {\scriptsize$\Delta_y m^z$};
    \node at (3.4,1.4) {\scriptsize$\Delta_y B^z$};
    \node at (-0.05,0.15) {\scriptsize{-1}};
    \node at (-0.05,1.2) {\scriptsize{0}};
    \node at (-0.05,2.2) {\scriptsize{$1$}};
    \node at (1,1) {\tiny{-5}};
    \node at (1.95,1) {\tiny{0}};
    \node at (2.8,1) {\tiny{5}};
    \node at (3.6,1) {\tiny{10}};
    \node at (0.2,2.6) {\footnotesize{\textbf{(d)}}};
    \def\x{2}
    \def\y{-1.4}
    \node[anchor=south west,inner sep=0] at (0.4+\x,1.2+\y) {\includegraphics[width=0.075\textwidth]{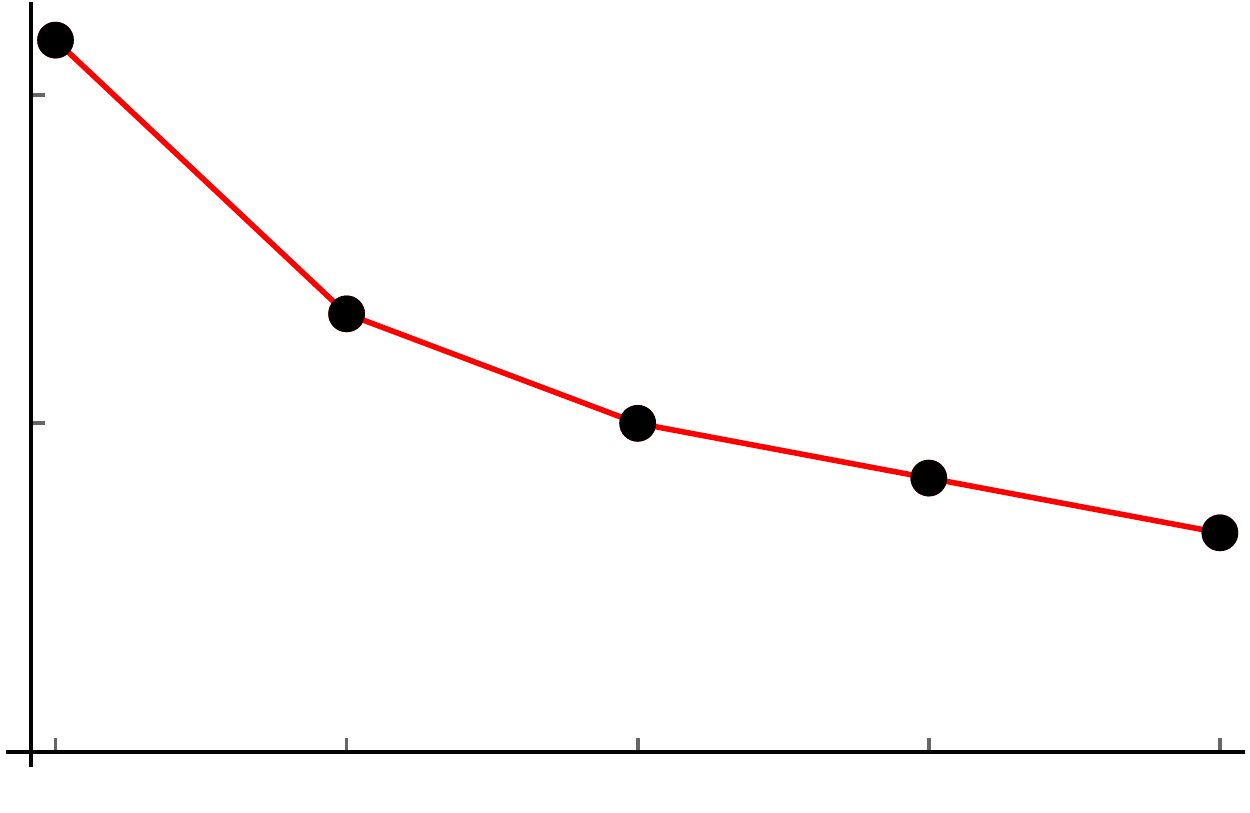}};
    \draw (0.43+\x,1.27+\y) -- (1.75+\x,1.27+\y) -- (1.75+\x,2.1+\y) -- (0.43+\x,2.1+\y) -- (0.43+\x,1.27+\y);
    \node at (0.52+\x,1.88+\y) {\tiny{4}};
    \node at (0.77+\x,1.63+\y) {\tiny{6}};
    \node at (1.07+\x,1.5+\y) {\tiny{8}};
    \node at (1.4+\x,1.43+\y) {\tiny{10}};
    \node at (1.63+\x,1.64+\y) {\tiny{12}};
    
    \node at (0.35+\x,1.97+\y) {\tiny{6}};
    \draw (0.44+\x,1.97+\y) -- (0.47+\x,1.97+\y);
    \node at (0.35+\x,1.62+\y) {\tiny{3}};
    \draw (0.44+\x,1.62+\y) -- (0.47+\x,1.62+\y);
    \node at (0.35+\x,1.3+\y) {\tiny{0}};
    
    \node [rotate=90] at (0.0+\x,1.8+\y) {\tiny{0 plateau}};
    \node [rotate=90] at (0.15+\x,1.6+\y) {\tiny{width}};
    
\end{tikzpicture}

\caption{Magnetization \textbf{(a, b)} and magnetization gradient \textbf{(c, d)} responses of a two-leg spin-1/2 ladder to an applied external magnetic field $B^z$ and magnetic field gradient $\Delta_y B^z$ respectively. In plots \textbf{(a)} and \textbf{(c)}, the two-leg ladder is coupled via spin-anisotropic XY interactions that do not preserve subsystem symmetry, while the ladder in plots \textbf{(b)} and \textbf{(d)} is coupled via ring-exchange interactions that have subsystem symmetry. Additionally, for we include the magnetization response to a constant $B^z$ of a chain that does not respect \emph{global} $U(1)$ $S^z$ rotation symmetry, which is depicted as a dashed red line in \textbf{(a)}. We superimposed numerical data for ladders with 4, 6, 8, 10, and 12 rungs. We clearly see that both magnetization and its gradient experience plateaus for the ring-exchange Hamiltonian, while there are no plateaus of magnetization gradient in the data for the XY-coupled ladder. In the magnetization gradient data for the ring-exchange model we see a small plateau at $\Delta_y m^z=0$ that monotonically shrinks with increasing system size as shown by the inset plot. It is possible that this plateau will not survive the thermodynamic limit.}
\label{fig:plateaus-noplateaus}
\end{figure}

We corroborate our results using the numerics presented in Fig. \ref{fig:plateaus-noplateaus}. In this figure we compare the magnetic responses of two types of spin-1/2 ladders. In Fig. \ref{fig:plateaus-noplateaus}a,c we show results for a two-leg spin-ladder with nearest neighbor XY couplings, while in Fig. \ref{fig:plateaus-noplateaus}b,d we show results for a two-leg spin ladder with ring-exchange terms (c.f. Eq. \ref{eqn:ham_bos_ring-exchange}). For each system we first show their response to a uniform magnetic field. For the XY ladder (Fig. \ref{fig:plateaus-noplateaus}a) we see magnetization plateaus at $M^z=0, \pm 1/2$ which matches the expected results since $2(S-M^z)\in\mathbb{Z}$ is the condition for a plateau for a two-leg system. We also find that the ring exchange model exhibits magnetization plateaus at $M^z=0,\pm 1/2$ (see Fig. \ref{fig:plateaus-noplateaus}b), in accordance to our expectations, since, as was mentioned above, for a two-leg ladder we must have $2(S-M^z)\in\mathbb{Z}$. In Fig. \ref{fig:plateaus-noplateaus}a we also overlayed a dashed red line showing the magnetization response of a single, nearest-neighbor spin chain coupled via $(2S^x_iS^x_{i+1}+S^y_iS^y_{i+1})$ interactions. These interactions  explicitly break the axial $U(1)$ symmetry corresponding to spin rotations around the $\hat{z}$ axis, and hence the system does not exhibit discrete magnetization plateaus, but instead smoothly interpolates between the two configurations where the average magnetization saturates. We chose to compare the XY ladder with an XY spin chain having broken $U(1)$ spin symmetry to make an analogy to the comparison between Figs. \ref{fig:plateaus-noplateaus}c,d between the XY ladder and ring-exchange ladder, where the former has broken $U(1)$ subsystem symmetry. 

Now we want to examine the magnetization gradient response of the XY ladder and the ring-exchange ladder. Let us consider applying a magnetization gradient centered around zero uniform applied magnetic field, i.e., we apply a $B^z=+h$ to the top leg of the ladder and $B^z=-h$ to the bottom one. The total magnetization of both systems stays at a magnetization plateau with $M^z=0$. For the XY ladder that respects only a global axial $U(1)$ symmetry, but not a subsystem $U(1)$, we see a smooth interpolation of the magnetization \emph{gradient} between the saturation points at $\Delta_y m^z=-1$ and $\Delta_y m^z=+1$. This is quite similar to the behavior of the \emph{magnetization} of the XY spin chain shown in Fig. \ref{fig:plateaus-noplateaus}a that does not respect the global axial $U(1)$ symmetry.

For the ring exchange model, which respects axial subsystem $U(1)$ symmetry, we find that $\Delta_y m^z$ exhibits a series of plateaus. The two most stable ones are located exactly at $\Delta_y m^z=\pm 1$ as we expect from Eq. \ref{eq:MGradDef}. In our numerical simulations, we also see a plateau at $\Delta_y m^z=0$, however, as we show in the inset plot in Fig. \ref{fig:plateaus-noplateaus}d, this plateau is shrinking rapidly as we increase the system size, and it is not clear if it will survive or not in the thermodynamic limit.

\subsubsection{Ising-Coupled Spin Ladder}
As a  brief aside, we can further illustrate the physics of magnetization gradient plateaus in one-dimensional ladders by connecting them to the ordinary magnetization plateaus in an effective single spin chain. To see this, consider a spin-1/2 ladder of length $L_x$ in a magnetic field with ring-exchange, Ising, and Zeeman couplings:
\begin{equation}
\begin{split}
H =  \sum_{x} [(J_{\square}S^+_{x,\uparrow} S^-_{x+1,\uparrow} S^+_{x+1,\downarrow} S^-_{x,\downarrow} + h.c.) + \lambda  S^z_{x,\uparrow} S^z_{x,\downarrow}\\ + h_\uparrow S^z_{x,\uparrow} + h_\downarrow S^z_{x,\downarrow}],
\label{eq:XYLadder}
\end{split}
\end{equation}
where ${\bf {S}}_{x,\uparrow/\downarrow}$ is the spin$-1/2$ operator on rung $x$ and the top ($\uparrow$) or bottom ($\downarrow$) leg. This Hamiltonian commutes with the total magnetization operator of the entire ladder ($\sum_x [S_{x,\uparrow}+ S_{x,\downarrow}]$), and the individual magnetization operators of each leg ($\sum_x S_{x,\uparrow }$ and $\sum_x S_{x,\downarrow }$). The former is a $U(1)$ global symmetry, while the latter are a pair of $U(1)$ subsystem symmetries. 

Here, we will be interested in the limit $\lambda \gg J_{\square}, |h_{\uparrow/\downarrow}|$, where every rung on the ladder will be pinned such that $\langle  S^z_{x,\uparrow} S^z_{x,\downarrow} \rangle = -\frac{1}{4}$. In this limit, the total magnetization of the system is fixed to be an integer $M^z = 0$, and the system is on a magnetization plateau. There are two configurations that satisfy this constraint: $\langle S^z_{x,\uparrow} \rangle =  \pm \frac{1}{2}$, $\langle S^z_{x,\downarrow} \rangle =  \mp \frac{1}{2}$.  Let us define a new effective spin degree of freedom on each rung, $\tilde{S}_r$ such that $\tilde{S}^z_x = S^z_{x,\uparrow} = -S^z_{x,\downarrow}$. Using this, we can also define: $\tilde{S}^+_x = S^+_{x,\uparrow} S^-_{x+1,\uparrow}$, $\tilde{S}^-_x = S^-_{x,\downarrow} S^+_{x+1,\downarrow}$. Combined with $\tilde{S}^z_x$, these operators satisfy the spin-1/2 algebra, and the spin ladder becomes:
\begin{equation}
\begin{split}
H = J_{\square} \sum_{x}\tilde{S}^+_x\tilde{S}^-_{x+1} + (h_\uparrow-h_\downarrow) \tilde{S}^z_x,
\end{split}
\label{eq:XYChain}
\end{equation}
which is the Hamiltonian for a single $XY$ spin \textit{chain} in an effective magnetic field given by $h_\uparrow-h_\downarrow$. 

Now let us consider the response of this system to external magnetic fields. When this system is placed in a uniform physical magnetic field ($h_1 = h_2$), the effective magnetic field vanishes so the system does not develop a magnetization as might be expected. However, if we instead consider a physical magnetic field gradient parallel to the ladder rungs (e.g., $h_1 = - h_2$), the effective magnetic field is non-vanishing and the system can develop an effective magnetization. The key point is that the magnetization of the effective spins in Eq. \ref{eq:XYChain} is equal to half the magnetization gradient of the original spin ladder: \begin{equation}
    \frac{1}{L_x}\sum_x \tilde{S}^z_x =\frac{1}{2}\frac{1}{L_x}\sum_x [S^z_{x,1} - S^z_{x,2}].
    \end{equation}
In conclusion, the effective magnetic field and magnetization associated to Eq. \ref{eq:XYChain} are respectively the physical magnetic field gradient and magnetization gradient of the spin ladder Eq. \ref{eq:XYLadder}. A magnetization plateau for the effective spins in Eq. \ref{eq:XYChain}, is thereby equivalent to a  magnetization gradient plateau for the physical spin in Eq. \ref{eq:XYLadder}. 

It is well known that the $XY$ spin chain has magnetization plateaus when the magnetization is equal to $\pm \frac{1}{2}$\cite{lieb1961}. From this, we can conclude that that the spin ladder Eq. \ref{eq:XYLadder} is at a plateau in the gradient of its magnetization when $\Delta_y m^z = \sum_x (S^z_{x,1} - S^z_{x,2})/L_x = 2 \sum_x \tilde{S}^z_x/L_x = \pm 1$. Since this model is also at an ordinary magnetization plateau at $M^z = 0$, this result agrees with Eq. \ref{eq:MGradDef}. It is worth noting that this result is only true in the limit $\lambda \gg |h_{\uparrow/\downarrow}|$. In the opposite limit, $|h_{\uparrow/\downarrow}| \gg \lambda, J$, the system will be at ordinary magnetic plateaus where the spins are aligned parallel to the magnetic field.

\subsection{Two-dimensional Spin Systems}
Now let us consider two-dimensional spin models. We will first derive the spin analog of the dipole LSM theorem from  Sec. \ref{sec:dip_LSM_A}. Working on a square periodic $L\times L=N a\times N a$ lattice, we will consider spin Hamiltonians which possess global $U(1)$ symmetry that acts by rotating all spins around the $\hat{z}$-axis by the same amount. The corresponding conserved quantity is the total magnetization $M^z$ of the system. 
We will additionally impose conservation of two components of the magnetic quadrupole moment $Q^M_{xz}$ and $Q^M_{yz}$ which is the analog of the conservation of the $x$ and $y$ components of the dipole moment for particles whose charge under the global $U(1)$ symmetry is itself a magnetic dipole moment pointing in the $z$-direction.

This setup is entirely analogous to the one considered in Sec. \ref{sec:dip_LSM_A}. 
It is natural then to consider Hamiltonians where the lowest-order dynamical terms are built of bosonic spin ring-exchange terms, as in (\ref{eqn:ham_bos_ring-exchange}).
Such systems were recently discussed in the literature\cite{you2019} where it was shown, that they naturally couple to the background symmetric rank-2 gauge field $A_{xy}$ with a Peierls phase factor.
To derive an LSM-type theorem we will briefly recount the argument already discussed in detail in the context of  dipole-conserving systems in Sec. \ref{sec:dip_LSM_A}.
We start by adiabatically driving the value of the background field $A_{xy}$ from $0$ to $2\pi/Na^2$ over time period $T$. 
This evolves the ground state of the system from $\ket{\Psi(0)}$ to $\ket{\Psi(T)}$. 
As this process is performed uniformly across the lattice, without breaking translational symmetry, $\ket{\Psi(T)}$ will remain an eigenstate of $T_x$ and $T_y$, provided that the translational invariance of the initial Hamiltonian was not spontaneously broken in its ground state $\ket{\Psi(0)}$.
Then, we apply the unitary twist operator
\begin{equation}
    U_{XY}=\exp\left(\frac{2\pi i}{aL}\sum_{\textbf{r}}xy\ S^z_{\textbf{r}}\right)
\end{equation} 
which removes the change in $A_{xy}$ and brings the Hamiltonian back to its original form. The resulting eigenstate $U_{XY}^{-1}\ket{\Psi(T)}$ has an energy infinitesimally close to the ground state in the thermodynamic limit, and it may be different from the original ground state $\ket{\Psi(0)}$.
If we consider the commutation relation between translations in the $\hat{x}$-direction and the twist operator, we obtain an additional phase factor:
\begin{equation}
    T_x U^{-1}_{XY} T^{-1}_{x}=U^{-1}_{XY}\text{e}^{\frac{2\pi i }{L}\sum_{\textbf{r}}y S^z_{\textbf{r}}}e^{-2\pi i\sum^{N}_{y=1} y S^z_{x=1,y}}
    \label{eqn:spin_extra_phase}
\end{equation}
where the first extra factor on the RHS of the equation contains the total magnetic quadrupole moment $Q^{M,tot}_{yz}$ of the system in the exponential. The phase in the second factor can take either integer or half-integer multiples of $2\pi$ depending on whether the spin $S$ is integer or half-integer, and whether the value of $N(N+1)/2$ is even or odd. 
For eigenvalues of the translation operator $T_x$ to be the same for $U_{XY}^{-1}\ket{\Psi(T)}$ and $\ket{\Psi(0)}$ the extra phase factors appearing in (\ref{eqn:spin_extra_phase}) must be trivial.
This yields the following condition for the uniqueness of the ground state:
\begin{equation}
    \begin{split}
    \frac{N(N+1)}{2}S+\sum_{\textbf{r}}\frac{y S^z_{\textbf{r}}}{Na} \in\mathbb{Z},
    \end{split}
    \label{eqn:spin_LSM}
\end{equation} which is very similar to the condition obtained in Sec. \ref{sec:dip_LSM_A}. For instance, consider integer $S$. The condition (\ref{eqn:spin_LSM}) then requires that, for the ground state to be unique, the total magnetic quadrupole moment $Q^M_{yz}$ must be an integer. We can repeat this calculation using translation in the $\hat{y}$-direction to derive
\begin{equation}
    \begin{split}
    \frac{N(N+1)}{2}S+\sum_{\textbf{r}}\frac{x S^z_{\textbf{r}}}{Na} \in\mathbb{Z},
    \end{split}
    \label{eqn:spin_LSM2}
\end{equation}\noindent which gives a similar condition but for $Q^M_{xz}.$

Now, let us move on to spin systems with $U(1)$ subsystem symmetry where the $S^z$ spin component is conserved on rows and columns of the lattice. We can again consider a twist operator that acts along a single column of spins with fixed coordinate $x=x_0$:
\begin{equation}
    U_{Y,x_0}=\exp\left(\frac{2\pi i}{L_y}\sum_{y}y S^{z}_{x_0,y}\right).
\end{equation}
For a periodic lattice and a subsystem symmetric Hamiltonian built from local ring-exchange terms such as Eq. (\ref{eqn:ham_bos_ring-exchange}), we can compare the energy of a twisted state with the original ground state $\ket{\Psi_0}.$ If we assume that $\ket{\Psi_0}$ does not spontaneously break the translational invariance and preserves a reflection symmetry $\hat{M}_y: y\to -y$, similar to (\ref{eqn:ham_difference}) we have:
\begin{equation}
    \langle\Psi_0 | U_{Y,x_0}^{-1} H U_{Y,x_0} - H |\Psi_0\rangle=O\left(\frac{1}{N_y}\right).
\end{equation}
Therefore, in the thermodynamic limit, where $N_y\to\infty$, the state $\ket{\tilde{\Psi}_0}=U_{Y,x_0}\ket{\Psi_0}$ has either exactly the same energy as the ground state $\ket{\Psi_0}$, or it is an excited state with an energy infinitesimally close to the ground state. 

We now want to see if $\ket{\Psi_0}$ and $\ket{\tilde{\Psi}_0}$ are orthogonal.
Assuming that the ground state does not spontaneously break the translational symmetry, i.e.,  $T_x\ket{\Psi_0}=\text{e}^{iP_{x0}}\ket{\Psi_0}$, we can show that the translation eigenvalue for $\ket{\tilde{\Psi}_0}$ may take a dstinct value:
\begin{equation}
\begin{split}
    T_x\ket{\tilde{\Psi}_0}&=T_x U_{Y,x_0} T^{-1}_{x}T_x\ket{\Psi_0}\\
    &=\text{e}^{iP_{x0}+\frac{2\pi i}{L_y}\sum_{y} y(S^z_{x_0+1,y}-S^z_{x_0,y})}\ket{\tilde{\Psi}_0}.
\end{split}
\end{equation}
Similar to the subsystem polarization introduced in Sec. \ref{sec:dip_LSM_A}, we introduce an analogous notion for spin systems -- a subsystem quadrupole polarization:
\begin{equation}
    \mathcal{Q}^M_{j z}(\mathfrak{s})=\frac{1}{2\pi}\text{Im}\log\langle\Psi_0|U_{j,\mathfrak{s}}|\Psi_0\rangle,
    \label{eqn:subsystem_qpol}
\end{equation}
where
\begin{equation}
    U_{j,\mathfrak{s}}=\exp\left(\frac{2\pi i}{L_j}\sum_{\textbf{r}\in \mathfrak{s}}x_j S^z_{\textbf{r}}\right),
\end{equation}
and where $j=x,y$.
Therefore, for the ground state to be unique we must have that the pair of magnetic quadrupole moments $\mathcal{Q}^M_{yz}$ computed along neighboring subsystems must differ by an integer number:
\begin{equation}
\begin{split}
    \text{e}^{\frac{2\pi i}{L_y}\sum_{y}y(S^z_{x_0+1,y}-S^z_{x_0,y})}\ket{\tilde{\Psi}_0}&=\ket{\tilde{\Psi}_0}\\
    \Leftrightarrow \mathcal{Q}^M_{yz}(x=x_0+1)&-\mathcal{Q}^M_{yz}(x=x_0)\in\mathbb{Z},
    \end{split}
    \label{eqn:spin_subsystem_rel}
\end{equation}
where $\mathcal{Q}^M_{yz}(x=x_0)$ is the magnetic quadrupolar polarization (\ref{eqn:subsystem_qpol}) along the column with fixed coordinate $x=x_0$.
Therefore, for the states $\ket{\Psi_0}$ and $\ket{\tilde{\Psi}_0}$ to have the same eigenvalue of the translation operator $T_x$ we need to require that the difference between the subsystem magnetic $\mathcal{Q}^M_{yz}$ quadrupole moments computed along two the adjacent rows of spins is an integer number. In general, since we can translate by any number of lattice constants in the $x$-direction, each column must have $\mathcal{Q}^M_{yz}$ that differ at most by an integer if we want to preserve translation symmetry and have a unique ground state. Noting that the total magnetic quadrupole moment $\mathcal{Q}^M_{yz}$ should satisfy the Eq. \ref{eqn:spin_LSM2} we can add in the relationship between subsystem quadrupole moments (\ref{eqn:spin_subsystem_rel}) to see that on a $N\times N$ lattice subsystem quadrupolarization must take the following set of values:
\begin{equation}
    \mathcal{Q}^M_{yz}(x=x_0)=\frac{n}{N}-\frac{N+1}{2}S,\ n\in\mathbb{Z}
\end{equation}
and similarly for $\mathcal{Q}^M_{xz}(y=y_0)$ on every column with fixed coordinate $y=y_0$.

Now, considering translations along $\hat{y}$ we find:
\begin{equation}
    T_y\ket{\tilde{\Psi}_0}=
    \text{e}^{iP_{y0}+2\pi i S^z_{x_0,1}-\frac{2\pi i}{N_y}\sum_{y=1}^{N_y} S^z_{x_0,y}}\ket{\tilde{\Psi}_0},
\end{equation}
which means that for the ground state $\ket{\Psi_0}$ to be unique we need the average magnetization $m^z_{x=x_0}$ of a single column at $x=x_0$ to satisfy: $(S-m^z_{x=x_0})\in\mathbb{Z}$, with $S$ being the total spin per unit cell of a subsystem. Hence, the states $\ket{\Psi_0}$ and $\ket{\tilde{\Psi}_0}$ are orthogonal unless the average magnetization of each subsystem $\mathfrak{s}$ satisfies
\begin{equation}
    S-m^z_{\mathfrak{s}}\in\mathbb{Z}.
\end{equation}

The physical consequences of these results are more subtle than the ladder case. We have found that in order for a system with magnetic quadrupole conservation to have a unique ground state the spin and magnetic quadrupolarization must satisfy an integer constraint. Furthermore, if the system has subsystem spin-rotation symmetry then each subsystem has to be on a magnetization plateau for the ground state to be unique. Thus, in the latter case, if we apply a spatially varying magnetic field that is constant along a family of subsystems, and weak enough not to drive any subsystem off its plateau, then the system will have a constant magnetization plateau response even to a \emph{spatially varying} magnetic field. For example, if we have subsystem symmetry in 2D along rows and columns parallel to $x$ and $y$ respectively, then our system will have a non-varying response to magnetic fields having only $x$ or only $y$ dependence as long as the field applied to any given subsystem is not strong enough to drive it off its plateau. In the former case without subsystem symmetry the system can exhibit a plateau in the magnetic quadrupolarization in the presence of a pure magnetic field gradient, i.e., a non-uniform magnetic field configuration that can have, at most, linear dependence on the spatial coordinates. Both of these possibilities suggest that for magnetic systems tuned to magnetization plateaus there can be a refinement of the magnetic response characterization based on how the system responds to non-uniform fields. Indeed, the systems we studied here can exhibit additional types of magnetic response plateaus when they have unique ground states.

\section{Luttinger-like theorem for dipoles}\label{sec:luttinger}
The LSM theorem has been used in non-perturbative arguments supporting Luttinger's theorem\cite{oshikawa1997,oshikawa2000B}. At the heart of these arguments is the connection between the momentum of a low-energy excitation and the particle filling. For a Fermi-liquid this relates the Fermi-surface, where low-energy excitations are created (having momentum of order $k_F$), to the electron filling, even in an interacting system. In this section we will apply similar arguments to show that some systems having $U(1)$ particle number and dipole conservation can support low-energy excitations having momentum determined by the filling of dipoles. Here we will just provide an example, and leave a full discussion, and generalization to higher dimensions to future work.

Let us consider a two-leg fermion ladder model parallel to the $x$-direction (let the lattice constant $a=1$). We will use the Hamiltonian\cite{dubinkinmaymannhughes2019} 
\begin{equation}
    H=\frac{J}{2}\sum_{i=1}^N(d^{\dagger}_i d^{\phantom{\dagger}}_{i+1}+h.c.)+U\sum_{i=1}^N n_{i\uparrow}n_{i\downarrow},
    \label{eqn:dipole_metal}
\end{equation}
where $\uparrow/\downarrow$ label the two legs of the ladder, $d_i\equiv c^\dagger_{i\downarrow}c_{i\uparrow}$ is a dipole annihilation operator for a dipole parallel to $y$, i.e., along the rungs of the ladder, $c^{\dagger}_{i\downarrow/\uparrow}$ is a fermion creation operator on the lower/upper legs respectively at site $i$, and the $n_{i\downarrow/\uparrow}$ are the fermion density operators on each leg. In Ref. \onlinecite{dubinkinmaymannhughes2019}, it was shown that when this system is at half-filling ($N_F=N$), and $U\gg J,$ then the dipole operators effectively become hard-core bosons having onsite anticommutation relations \begin{equation}
    \{d^\dagger_i,d_i\}=1,\ \{d^\dagger_i,d^\dagger_i\}=\{d_i,d_i\}=0,
    \label{eqn:dipole_anticommute2}
\end{equation} and commuting off site. Thus, in this limit this model becomes a hopping model for $y$-oriented dipoles that behave as hardcore bosons. We can identify up-dipoles (down-dipoles) with a configuration where, at a particular unit cell $i$ there is a fermion on the upper leg (lower leg) and no fermion on the lower leg (upper leg). Based on this, we can define the total dipole number operator as $N_D = \sum^N_{i=1} [n_{i\uparrow} - n_{i\downarrow}]$, and the $y$ component of the polarization as $p^y = N_D/N$. It is clear that the Hamiltonian in Eq. \ref{eqn:dipole_metal} conserves the dipole number $N_D$. Since the total fermion number ($N_F = \sum^N_{i=1} [n_{i\uparrow} + n_{i\downarrow}]$) is also conserved, the fermion number on each leg of the ladder ($N_{\uparrow}=\sum^N_{i=1} n_{i\uparrow}$ and  $N_{\downarrow}=\sum^N_{i=1}n_{i\downarrow}$) is conserved as well. 

To prove a Luttinger-like theorem we want to show that the low-energy modes of this model at some filling of dipoles have momentum related to that dipole filling.  Let us take the ground state of the system to be $\ket{\Psi_0}$. We will consider the twisted variational state $\ket{\tilde{\Psi}_0} = \exp(2\pi i\sum^N_{j =1}j n_{j,\uparrow}/N)\ket{\Psi_0}$. A calculation analogous to what we have presented in detail above for the ring exchange model shows that the energy of this state is within $\mathcal{O}(1/N)$ of the ground state energy. We can now calculate the momentum of this state. If $\ket{\Psi_0}$ is an eigenstate of the lattice translation operator $T_x$ with eigenvalue $e^{i P_{x0}}$, then $\ket{\tilde{\Psi}_0}$ will have an eigenvalue $e^{iP_{x0} + 2\pi i\sum^N_{i =1} n_{i,\uparrow}/N}$. Using the relation $\sum^N_{i =1} n_{i,\uparrow}/N = \frac{1}{2N} (N_D + N_F)$, (and also that $N_F=N$ since the fermions are half-filled), there must be a low energy mode with momentum $[P_{x0} + \pi(p^y + 1)]$, where we recall $p_y$ is the $y$-component of the charge polarization. Similarly, if we twist the ground state with the inverse of the operator above we will find  another low energy mode with momentum $[P_{x0} - \pi(p^y + 1)]$. We recall that these modes are only guaranteed to be orthogonal to the untwisted ground state if the polarization $p^y$ is not an integer.  We will argue below that these points form an analog of a Fermi surface for dipoles with Fermi wavevector
\begin{equation}
    k^{(dipole)}_F=\pi\left(p^y+1\right).
    \label{eqn:Luttinger_dipole}
\end{equation} 

Alternatively, we can derive these results with an explicit solution of this model. If the dipoles are effectively hard-core bosons, this model can be transformed into a spin-1/2 XY model using:
\begin{equation}
    S^\alpha_i=\frac12 \vec{c}_{i}^{\ \dagger} \sigma^\alpha \vec{c}_{i},\ \text{where}\  \vec{c}_{i}=(c_{i,\uparrow},c_{i\downarrow})^T,
    \label{eqn:fermions_to_spin}
\end{equation}
so that
\begin{equation}
    d^\dagger_i=2S^+_i,\ d_i=2S^-_i.
    \label{eqn:dipoles_to_spin}
\end{equation}
The resulting spin Hamiltonian is
\begin{equation}
    H=2J\sum_{i=1}^N \left(S^+_iS^-_{i+1}+S^-_{i}S^+_{i+1}\right).
    \label{eqn:heisenberg_model}
\end{equation}
It is well-known that such an XY model is exactly solvable in 1D  via a Jordan-Wigner transformation:
\begin{equation}
    S^+_i=\text{e}^{i\pi\sum_{j=1}^{i-1}f^\dagger_j f_j}f^\dagger_i,\ S^-_i=\text{e}^{-i\pi\sum_{j=1}^{i-1}f^\dagger_j f_j}f_i,
    \label{eqn:jw-map}
\end{equation} and the resulting transformed Hamiltonian is:
\begin{equation}
    H=2J\left(\sum_{i=1}^{N-1}f^\dagger_i f_{i+1}+\text{e}^{i\pi\sum_{j=1}^{N}f^\dagger_j f_j}f^\dagger_N f_1\right)+h.c.
    \label{eqn:fermion_ham}
\end{equation}\noindent where $f_{i}$ is the annihilation operator for a Jordan-Wigner fermion on site $i.$

Using these mappings we can identify the low-energy excitations of the dipole model with the Fermi-surface excitations of the Jordan-Wigner fermions. These excitations occur at momentum $\pm k^{(dipole)}_{F}$ which is directly proportional to the density of Jordan-Wigner fermions, and through the mappings above to the density of $y$-dipoles. Precisely we have $2k^{(dipole)}_F=2\pi\nu$, where $\nu$ is the fraction of up-dipoles in the system. Alternatively we can rewrite dipole density as:
\begin{equation}
    \nu=\frac{N_\uparrow}{N_\uparrow+N_\downarrow}=\frac12\frac{N_\uparrow-N_\downarrow}{N_\uparrow+N_\downarrow}+\frac{1}{2}=\frac12(p^y+1), 
    \label{eqn:fractional_dipole-polarization}
\end{equation} where $p^y$ is the charge polarization in the $y$-direction. 
Thus we can relate the area enclosed by a Fermi surface in 1D to the polarization of the dipole chain in the transverse direction and we recover Eq. \ref{eqn:Luttinger_dipole}.

We expect that results like this can apply beyond one-dimensional ladders. As an example, we could consider a model for a 2D \emph{dipole metal} recently discussed in Ref. \onlinecite{dubinkinmaymannhughes2019}. 
This model is built by stacking dipole ladder models (which are parallel to $\hat{x}$) into the $y$-direction and introducing dipole hopping terms between the nearest-neighbor rungs of two neighboring ladders. Effectively, the model describes a system of free $y$-dipoles that can move across a rectangular lattice. 
It was shown that this model can be Jordan-Wigner transformed to a fermionic tight-binding model which has a well-defined Fermi surface. 
This transformation translates number operators for $y$-dipoles into ordinary number operators for the Jordan-Wigner fermions. Hence Luttinger's theorem for a two-dimensional fermionic model, when translated to a dipole language, once again relates an area enclosed by a Fermi surface to the density of $y$-dipoles in the lattice, or, in other words, to the $\hat{y}$ polarization of the ground state.

We note a possible connection to the recent work in Ref. \onlinecite{prem2017emergent}, where elementary dipole particles having a fixed dipole moment were considered. In comparison, however, the statistics of those particles was taken to be fermionic (in our case they are hard-core bosons), and the interactions between particles were taken into account.  It was then argued that this system develops a stable interacting Fermi liquid with a Fermi surface elongated in the direction of the dipole moment.
A Luttinger theorem for such fermionic dipoles, which we do not prove here, would also necessarily relate an area enclosed by a Fermi surface to the density of fermionic dipoles, i.e., the polarization of the system in the dipole moment direction.

\section{Conclusions}
In this paper we derived several non-perturbative results for dipole-conserving Hamiltonians and their spin counterparts. We provided a generalization of the LSM theorem to multipole-conserving systems, and find that for dipole conserving systems, a unique gapped, translationally invariant ground state is possible only if the bulk polarization is integer (integer filling of dipoles). A rational polarization of $p/q$ implies that there are at least $q$ degenerate ground states. Furthermore, if the system both conserves polarization and has a $U(1)$ charge conservation symmetry along subsystems, a unique gapped, translationally invariant ground state is possible only if the filling in each subystem is an integer. A rational filling implies either a gapless system or a ground state degeneracy. We also provided the spin counterpart of this theorem, that applies to spin systems that have conserved magnetic quadrupole moments and possibly preserve spin-rotation symmetry on subsystems. These systems can experience plateaus in the magnetic response in some types of non-uniform fields.  Finally we have also discussed a possible extension of a Luttinger-like theorem to dipole systems. 

From these results, we have been able to place strong constraints on the low energy physics of systems having conserved multipole moments. As with the famous results of Lieb, Schultz and Mattis, these results can be used to study strongly correlated systems, where normal perturbative methods fail. Much is still unknown about multipole conserving systems on lattices\cite{vijay2016, nandkishore2019, pai2019scar}, and in the continuum \cite{pretko2018,gromov2019,may2019}, and our results may prove useful in these contexts. These results also hint at possible exotic gapped phases that have fractional polarization/dipole moment, in analogy to topologically ordered systems having fractional charge. The new types of magnetic response plateaus we predicted may also belong to topological phases, analogous to the Haldane phase in SPT spin chains. Experimentally, our results can be tested in cold-atom systems, where dipole conserving systems can be constructed\cite{buchler2005, li2005, dai2017}. These cold-atom systems may be an interesting place to look for the aforementioned exotic phases. While we have focused primarily on 1D and 2D we expect the results can be extended straightforwardly to higher dimensions, and with a variety of conserved types of multipole moments. Finally, it could prove useful exploring possible connections between our LSM-type theorems and similar results recently acquired in the context of systems with higher-form symmetries\cite{Kobayashi19}.

{\bf{Note:}} During the preparation of this manuscript we became aware of a recent work titled ``Lieb-Schultz-Mattis type constraints on Fractonic Matter" by Huan He, Yizhi You, and Abhinav Prem\cite{yizhiLSM}. Our work has some overlapping concepts and results with this article, but both were carried out independently.

\acknowledgements
We thank Yizhi for useful discussions, and for pointing us to their recent article. OD and TLH acknowledge support from the US National Science Foundation under grant DMR 1351895-CAR, and the MRSEC program under  NSF  Award  Number  DMR-1720633  (SuperSEED)for support. JMM is supported by the National Science Foundation Graduate Research Fellowship Program under Grant No. DGE – 1746047.

\bibliography{References.bib}

\end{document}